\documentclass[12pt]{article}
\usepackage{epsf}
\def\sqr#1#2{{\vcenter{\vbox{\hrule height.#2pt
        \hbox{\vrule width.#2pt height#1pt \kern#1pt
           \vrule width.#2pt}
        \hrule height.#2pt}}}}

\def\lsim{{\displaystyle
{{\raise-8pt\hbox{$ <$}}
\atop{\raise5pt\hbox{$\sim$}}}}}
\def\gsim{{\displaystyle
{{\raise-8pt\hbox{$ >$}}
\atop{\raise5pt\hbox{$\sim$}}}}}
\def\limit#1#2{\smash { \mathop{#1} \limits_{#2} }  }
\def\slsim{{\displaystyle
{{\raise-8pt\hbox{$\scriptstyle <$}}
\atop{\raise5pt\hbox{$\scriptstyle \sim$}}}}}
\def\sgsim{{\displaystyle
{{\raise-8pt\hbox{$\scriptstyle  >$}}
\atop{\raise5pt\hbox{$\scriptstyle \sim$}}}}}
\newskip\humongous \humongous=0pt plus 1000pt minus 1000pt
\def\caja{\mathsurround=0pt}
\def\eqalign#1{\,\vcenter{\openup1\jot \caja
        \ialign{\strut \hfil$\displaystyle{##}$&$
        \displaystyle{{}##}$\hfil\crcr#1\crcr}}\,}

\newcommand{\sump}[0]{\sum_{(h,g)}\!{\raise 4pt \hbox{$'$}}\,}
\newcommand{\sumpf}[0]{\sum_{(H^{\rm f},G^{\rm f})}\! \! \! \! {\raise
4pt
\hbox{$'$}}\,}

\catcode`\@=11
\newcount\hour
\newcount\minute
\newtoks\amorpm
\hour=\time\divide\hour by60
\minute=\time{\multiply\hour by60 \global\advance\minute by-\hour}
\edef\standardtime{{\ifnum\hour<12 \global\amorpm={am}%
        \else\global\amorpm={pm}\advance\hour by-12 \fi
        \ifnum\hour=0 \hour=12 \fi
        \number\hour:\ifnum\minute<10 0\fi\number\minute\the\amorpm}}
\edef\militarytime{\number\hour:\ifnum\minute<10 0\fi\number\minute}
\def\draftlabel#1{{\@bsphack\if@filesw {\let\thepage\relax
   \xdef\@gtempa{\write\@auxout{\string
      \newlabel{#1}{{\@currentlabel}{\thepage}}}}}\@gtempa
   \if@nobreak \ifvmode\nobreak\fi\fi\fi\@esphack}
        \gdef\@eqnlabel{#1}}
\def\@eqnlabel{}
\def\@vacuum{}
\def\draftmarginnote#1{\marginpar{\raggedright\scriptsize\tt#1}}
\def\draft{\oddsidemargin -.2truein
        \def\@oddfoot{\sl preliminary draft \hfil
        \rm\thepage\hfil\sl\today\quad\militarytime}
        \let\@evenfoot\@oddfoot \overfullrule 3pt
        \let\label=\draftlabel
        \let\marginnote=\draftmarginnote
   \def\@eqnnum{(\theequation)\rlap{\kern\marginparsep\tt\@eqnlabel}%
\global\let\@eqnlabel\@vacuum}  }
\relax
\def\Trp{\,{\rm Tr'}\, }

\def\Im{\,{\rm Im}\, }

\def\bJ{\overline{J}}

\def\bP{\overline{P}}
\def\bQ{\overline{Q}}

\def\bE{\overline{E}}
\def\bH{\overline{H}}
\def\bF{\overline{F}}
\def\bPhi{\overline{\Phi}}

\def\bOmega{\overline{\Omega}}

\def\g{\gamma}
\def\thefootnote{\fnsymbol{footnote}}
\def\be{\begin{equation}}
\def\ee{\end{equation}}
\def\ba{\begin{eqnarray}}
\def\ea{\end{eqnarray}}
\def\bs{\begin{subequations}}
\def\es{\end{subequations}}

\def\th{\vartheta}

\def\l{\lambda}

\def\d{\delta}
\def\g{\gamma}
\def\t{\tau}

\def\t{\tau}
\def\im{\, {\rm Im}\, \tau}

\def\sp{\ , \ \ }
\def\ifd{\int_{\cal F}\frac{{\rm d}^2\tau}{\im}}
\newcommand{\ar}[2]{{#1\atopwithdelims[]#2}}

\def\ee{\end{equation}}
\def\bea{\begin{eqnarray}}
\def\eea{\end{eqnarray}}
\def\nn{\nonumber}

\def\np#1#2#3{Nucl. Phys. {\bf{B#1}} (#2) #3}
\def\pl#1#2#3{Phys. Lett. {\bf{B#1}} (#2) #3}

\newcommand{\uarrw}[0]{\mathrel{
{\raise.5ex\vbox{\hrule width 1cm}\hskip-6pt\rightarrow}}}
\newcommand{\underarrow}[1]{\mathop{\uarrw}_{#1}}
\def\thebibliography#1{%
\vskip 0.5cm \centerline{\bf References}
\list{%
[\arabic{enumi}]}{\settowidth\labelwidth{[#1]}
\leftmargin\labelwidth
\advance\leftmargin\labelsep
\usecounter{enumi}}
\def\newblock{\hskip .11em plus .33em minus .07em}
\sloppy\clubpenalty4000\widowpenalty4000
\sfcode`\.=1000\relax}

\renewcommand{\theequation}{\arabic{section}.\arabic{equation}}

\renewcommand{\section}{\setcounter{equation}{0}\@startsection%
{section}{1}{0mm}{-\baselineskip}{0.5\baselineskip}%
{\normalfont\normalsize\bfseries}}

\renewcommand{\subsection}{\@startsection%
{subsection}{2}{0mm}{-\baselineskip}{0.5\baselineskip}%
{\normalfont\normalsize\slshape}}
\begin{document}
\renewcommand{\theequation}{\arabic{section}.\arabic{equation}}
\begin{titlepage}
\begin{flushright}
CERN-TH/98-195 \\
CPTH-S609.0498 \\
LPTENS/98/15 \\
hep-th/9808024\\
\end{flushright}
\begin{centering}
\vspace{.15in}
\boldmath
{\bf NON-PERTURBATIVE GRAVITATIONAL CORRECTIONS IN A CLASS OF $N=2$
STRING DUALS}$^{\ \ddagger}$\\
\unboldmath
\vspace{.3in}
{Andrea GREGORI}$^{\ 1}$, {Costas KOUNNAS}$^{\ 1,\, \ast}$\\
\medskip
and \\
\medskip
{P. Marios PETROPOULOS}$^{\ 1,\, 2}$\\
\vspace{.1in}
{\it $^1 $ Theory Division, CERN}\\
{\it 1211 Geneva 23, Switzerland}\\
\medskip
{\it $^2 $ Centre de Physique Th\'eorique,
Ecole Polytechnique}$^{\ \diamond}$\\
{\it 91128 Palaiseau Cedex, France}\\
\vspace{.15in}
{\bf Abstract}\\
\end{centering}
\medskip
We investigate the non-perturbative equivalence of some
heterotic/type II dual pairs with $N=2$ supersymmetry. The
perturbative heterotic scalar manifolds are respectively
$ SU(1,$ $1)\big/U(1) \times  SO(2,2+N_V)\big/  SO(2)
\times SO(2+N_V)  $ and $ SO(4,4+N_H)\big/
SO(4)\times SO(4+N_H)$ for moduli in the vector multiplets
and hypermultiplets. The models under consideration correspond, on
the
type II side, to self-mirror Calabi--Yau threefolds with Hodge
numbers
$h^{1,1}= N_V +3=h^{2,1}=N_H +3$, which are $K3$ fibrations. We
consider three classes of dual pairs, with $N_V=N_H=8$, 4 and 2.
The models with $h^{1,1}=7$ and 5 provide new constructions, while
the
$h^{1,1}=11$, already studied in the literature, is reconsidered
here. Perturbative $R^2$-like corrections are computed on the
heterotic side by using a universal operator whose amplitude has no
singularities in the $(T,U)$ space, and can therefore be
compared with the type II side result.
We point out several properties connecting $K3$ fibrations and
spontaneous breaking of the $N=4$ supersymmetry to $N=2$. As a
consequence of the reduced $S$- and $T$- duality symmetries, the
instanton numbers in these three classes are restricted to integers,
which are multiples of 2, 2 and 4, for $N_V=8$, 4 and 2,
respectively.
\vspace{.15in}
\begin{flushleft}
CERN-TH/98-195 \\
CPTH-S609.0498 \\
LPTENS/98/15\\
August 1998 \\
\end{flushleft}
\hrule width 6.7cm
\vskip.1mm{\small \small \small
$^\ddagger$\ Research partially supported by the EEC under the contracts
TMR-ERBFMRX-CT96-0045 and TMR-ERBFMRX-CT96-0090.
\\
$^\ast$\ On leave from {\it Laboratoire de Physique Th\'eorique de
l'Ecole Normale Sup\'erieure,} \\
{\it CNRS,} 24 rue Lhomond, 75231 Paris Cedex 05, France.\\
$^\diamond$ Unit{\'e} propre du {\it Centre National de la Recherche
Scientifique}.}
\end{titlepage}
\newpage
\setcounter{footnote}{0}
\renewcommand{\thefootnote}{\arabic{footnote}}

\setcounter{section}{0}
\section{Introduction}

Different perturbative string theories with the same number of
supersymmetries might be equivalent at the non-perturbative level
\cite{ht,w1,pw}. There is a conjectured duality between the
heterotic string compactified on $T^4\times T^2$ and the type IIA
(IIB) string compactified on $K3\times T^2$ \cite{ht}. Both
theories have $N=4$ supersymmetry and 22 massless vector multiplets
in four dimensions. In both theories the space of the moduli-field
vacuum expectation values is spanned by 134 physical scalars, which
are coordinates of the coset space \cite{ht,fk}
\be
\left({SL(2,R)\over U(1)}\right)_S
\times\left({SO(6,6+r)\over SO(6)\times
SO(6+r)}\right)_{T}\, , \ \ r=16\, .
\ee
On the heterotic side, the dilaton $S_{\rm Het}=S$ is in the
gravitational multiplet, while on the type II side it is one of the
moduli of the vector multiplets: $S_{\rm II}=T^1$, where $T^1$ is the
volume form of the two-torus. Thus the duality
relation implies an interchange between the fields of $S$ and $T$
manifolds \cite{ht}: the perturbative heterotic states are mapped to
non-perturbative states of the type II theory and vice versa.
Consequently, perturbative $T$-duality of the type II strings implies
$S$-duality \cite{s} of the heterotic string and vice versa.

Several arguments support this duality conjecture.
For instance, the anomaly cancellation of the six-dimensional
heterotic string implies that there should be a one-loop correction
to the gravitational $R^2$ term in the type II theory. Such a term
was found by direct calculation in \cite{vw2}. Its one-loop threshold
correction upon compactification to four dimensions \cite{hmn=4}
implies instanton corrections on the heterotic side due to
five-branes wrapped around the six-torus.

Heterotic/type II dual pairs with lower rank (i.e. $r<16$)
and with $N=4$ supersymmetry, share properties
similar to those just mentioned. The study of
such theories and the determination of the heterotic non-perturbative
corrections to the $R^2$ terms were considered in \cite{6auth}. Here
we
would like to extend this analysis for $N=2$ heterotic/type II dual
theories.

In general (non-freely-acting) symmetric orbifolds still give rise to
$N=2$ heterotic/type II dual pairs in four dimensions
\cite{kv,fhsv,re}. On the heterotic side they can be interpreted as
$K3$ plus gauge-bundle compactifications, while on the type II side
they are Calabi--Yau compactifications of the ten-dimensional type
IIA theory. The heterotic dilaton is in a vector multiplet and the
vector moduli space receives both perturbative and non-perturbative
corrections. On the other hand, the hypermultiplet moduli space does
not receive perturbative corrections;
if $N=2$ is assumed unbroken, it receives no non-perturbative
corrections either. On the type II side the dilaton is in
a hypermultiplet and the prepotential for the vector multiplets
receives only tree-level contributions. The tree-level type II
prepotential was computed and shown to give the correct one-loop
heterotic result. This provides a quantitative test of the duality
\cite{kv,re} and allows us to reach the non-perturbative corrections
of
the heterotic side.

The purpose of this paper is to provide quantitative tests of $N=2$
heterotic/type II duality. Quantitative tests of non-perturbative
dualities can be obtained by considering the renormalization of
certain terms in the effective action of the massless fields.
Extended supersymmetry plays an essential role in this since it
allows the existence of BPS states that, being short representations
of the supersymmetry algebra, are (generically) non-perturbatively
stable and provide a reliable window into non-perturbative physics.
There are terms in the effective action, the couplings of which can
be shown to obtain contributions only from BPS states. The relevant
structures for this analysis are helicity supertrace formulae, which
distinguish between various BPS and non-BPS states
\cite{6auth,bk,kk,n=3}.
For ($N=2$)-supersymmetric models, these supertraces appear in
particular
in the two-derivative terms $R^2$ or in a special class of
higher-order terms constructed out of the Riemann tensor and the
graviphoton field strength \cite{fg}. In the four-dimensional
heterotic string, these terms are anomaly-related and it can be shown
that they receive only tree-level and one-loop corrections. In higher
dimensions, they receive no non-perturbative correction \cite{bk2}.

In this paper we investigate the non-perturbative equivalence of some
heterotic/type II dual pairs with $N=2$ supersymmetry. The
perturbative heterotic scalar manifolds are

\be
{SU(1,1)\over U(1)}\times {SO(2,2+N_V)\over SO(2)\times SO(2+N_V)}
\ \ {\rm and}
\ \ {SO(4,4+N_H)\over SO(4)\times SO(4+N_H)}
\ee
for moduli in the vector multiplets and hypermultiplets,
respectively \footnote{In the models under consideration, $N_V$ and
$N_H$ are the number of vector multiplets and hypermultiplets, 
respectively, which on
the type II construction are originated from the twisted sectors. The
total number of those are $N_V+2$ and $N_H+4$ for heterotic constructions,
$N_V+3$ and $N_H+3$ in the type II cases. These numbers neither include the 
vector-tensor multiplet present in heterotic models, which is dual
to a vector multiplet, nor the tensor multiplet of the type II ground 
states dual to a hypermultiplet.}. On
the type II side, the models under consideration correspond to
self-mirror Calabi--Yau threefolds with Hodge numbers $h^{1,1}= N_V +
3=h^{2,1}=N_H + 3$, which are $K3$ fibrations, necessary condition
for the existence of heterotic duals \cite{klm,al}.

In Section \ref{II} we examine the
$N=2$ type II models with $N_V=N_H=8$, 4, and 2, which are particular
examples of $Z_2 \times Z_2$
symmetric orbifolds. A detailed and complete classification of such
orbifolds will appear in \cite{gkpr}.
The models with $h^{1,1}=7$ and 5 provide new
constructions, while the one with $h^{1,1}=11$ was already considered
in the
literature \cite{fhsv,hmn=2}. It is reconsidered here since, as we
will
see, our choice of the $R^2$ terms is not exactly the one taken
previously.
We derive the perturbative type II $R^2$ corrections and
point out several properties that connect the $K3$
fibrations and the spontaneous breaking of the
$N=4$ to the $N=2$  supersymmetry.

In Section \ref{het} we construct the heterotic duals. The
construction  is performed in a unified formalism for all values of
$N_V$. Particular attention is payed to the singularities arising
along various lines in the $(T,U)$ moduli space, where extra massless
states appear. This analysis serves as a guideline in the
determination of an operator on the heterotic side that reproduces
the $R^2$ term already considered in the type II side. This is
achieved in Section \ref{hetthr}, where it is precisely shown that
indeed there exists a universal, holomorphic and modular-covariant
operator $Q^{\prime 2}_{\rm grav}$; its coupling constant
receives  perturbative corrections, regular at every point of  the
two-torus moduli space. This enables us to check the heterotic/type II
duality conjecture at the perturbative and non-perturbative level.

Our conclusion and comments are given in Section \ref{con}.

\section{Type II reduced-rank models and
gravitational corrections}\label{II}
\subsection{Construction of the reduced-rank models}

\noindent {\sl a) The $N_V=N_H=8$ model}

\noindent We start by considering the $N=2$ type II model with
$N_V=N_H=8$. This model is obtained by compactification of the
ten-dimensional superstring on a Calabi--Yau threefold with
$h^{1,1}=h^{2,1}=11$ \cite{fhsv}. It can be constructed in two steps.
We start with the type II $N=4$ supersymmetric model defined by
$K3\times T^2$ compactification. The massless spectrum of this model
contains the $N=4$ supergravity multiplet as well as 22 vector
multiplets. For convenience, we will go to the $T^4/Z_2$ orbifold
limit of $K3$, where it can easily be shown that the 6 graviphotons
and
6 of the vector multiplets are coming from the untwisted sector,
while the remaining 16 come from the twisted sector and are in
one-to-one
correspondence with the 16 fixed points of the
$Z_2$ action. We then break the $N=4$ supersymmetry to $N=2$ by
introducing an
extra $Z_2$ projection in which two of the $T^4$ coordinates are
shifted; the remaining two coordinates of $T^4$ together with the two
coordinates of $T^2$ are $Z_2$-twisted.

If we indicate by $Z_2^{(\rm o)}$ the projection that defines the
$T^4/Z_2^{(\rm o)}$ orbifold limit of $K3$, and by $Z_2^{(\rm f)}$
the second projection defined above, we can summarize the action of
the
two $Z_2$'s on the three complex (super-)coordinate planes as in
Table 1; $R$ indicates the twist, while $T$ is a half-unit lattice
shift.

\begin{center}
\begin{tabular}{| c | c | c | c |}
\hline
Orbifold & Plane 1 & Plane 2 & Plane 3 \\ \hline
$Z_2^{(\rm o)}$ & 1 & $R$ & $R$ \\ \hline
$Z_2^{(\rm f)}$ & $R$ & $T$ & $RT$ \\ \hline
$Z_2^{(\rm o)} \times Z_2^{(\rm f)}$ & $R$ & $RT$ & $T$ \\ \hline
\end {tabular}
\end{center}

\centerline{Table 1. The action of $Z_2^{(\rm o)}
\times Z_2^{(\rm f)}$ on the $T^6$.}

\noindent
Since the translations on the second and third
planes are non-vanishing, the $Z_2^{(\rm f)}$-operation has no fixed
points
and there are therefore no extra massless states coming from the
twisted sectors; the massless spectrum of this model contains the
$N=2$ supergravity multiplet, 11 vector multiplets
(3 from the untwisted and 8 from the twisted sector), 1 tensor
multiplet and 11 hypermultiplets (3 are from the untwisted sector and
8 from the twisted sector). The tensor multiplet is the type II dilaton
supermultiplet, and it is equivalent to an extra hypermultiplet.

The partition function of the model reads:
\ba
Z_{\rm II}^{N_V}&=&
{1 \over \Im \tau \, \vert\eta\vert^{24}}
{1\over 4}\sum_{H^{\rm o},G^{\rm o}}\sum_{H^{\rm f},G^{\rm f}}
\Gamma_{6,6}^{N_V}\ar{H^{\rm o},H^{\rm f}}{G^{\rm o},G^{\rm f}}
\nonumber\\
&&\times {1\over 2}\sum_{a,b}
(-)^{a+b+ab}
\vartheta \ar{a}{b} \,
\vartheta \ar{a+H^{\rm o}}{b+G^{\rm o}}\,
\vartheta \ar{a+H^{\rm f}}{b+G^{\rm f}} \,
\vartheta \ar{a-H^{\rm o}-H^{\rm f}}{b-G^{\rm o}-G^{\rm f}}
\nonumber\\
&&\times
{1\over 2}\sum_{\bar a,\bar b}
(-)^{\bar{a}+\bar{b}+\bar{a}\bar{b}}
\bar{\vartheta} \ar{\bar{a}}{\bar{b}}\,
\bar{\vartheta} \ar{\bar{a}+H^{\rm o}}{\bar{b}+G^{\rm o}}\,
\bar{\vartheta} \ar{\bar{a}+H^{\rm f}}{\bar{b}+G^{\rm f}}\,
\bar{\vartheta} \ar{\bar{a}-H^{\rm o}-H^{\rm f}}{\bar{b}-G^{\rm
o}-G^{\rm f}}\, ,
\label{zII}
\ea
with
\be
\Gamma_{6,6}^{N_V=8}\ar{H^{\rm o},H^{\rm f}}{G^{\rm o},G^{\rm f}}=
\Gamma_{2,2}^{(1)} \ar{H^{\rm f}\vert 0}{G^{\rm f}\vert 0}
\,
\Gamma_{2,2}^{(2)} \ar{H^{\rm o}\vert H^{\rm f}}{G^{\rm o}\vert
G^{\rm
f}}
\,
\Gamma_{2,2}^{(3)} \ar{H^{\rm o}+H^{\rm f}\vert H^{\rm f}}{G^{\rm
o}+G^{\rm f}\vert G^{\rm f}}\, ,
\label{z6688}
\ee
the contribution of the six compactified left
and right coordinates $X^I$ and $\bar X^I$,
where $(H^{\rm o},G^{\rm o})$ refer to the boundary conditions
introduced by the
projection $Z_2^{(\rm o)}$ and $(H^{\rm f},G^{\rm f})$ to the
projection
$Z_2^{(\rm f)}$. Here we have introduced the twisted and shifted
characters of a $c=(2,2)$ block,
$\Gamma_{2,2} \ar{h\vert h'}{g\vert g'}$; the first column refers to
the twist, the second to the shift. The non-vanishing components
are the following:
\ba
\Gamma_{2,2} \ar{h\vert h'}{g\vert g'}
&=&
{4\,  \vert\eta \vert^6\over
\left\vert
\vartheta{1+h\atopwithdelims[] 1+g}\,
\vartheta{1-h\atopwithdelims[] 1-g}
\right\vert} \ , \ \
{\rm for\ } (h',g')=(0,0) \ \
{\rm or\ } (h',g')=(h,g) \nn \\
&=&
\Gamma_{2,2} \ar{h'}{g'}\ , \ \
{\rm for\ } (h,g)=(0,0)\, ,
\label{g22ts}
\ea
where $\Gamma_{2,2} \ar{h'}{g'}$ is the $Z_2$ shifted $(2,2)$ lattice
sum. The shift has to be specified by
the way it acts on the windings and momenta (such technical details
can be found in various references, our conventions are those of
\cite{6auth}). All the moduli of the type II compactification at hand are
contained in expression (\ref{z6688}), which depends on the volume forms 
$T^1, T^2, T^3 $ and the complex-structure forms
$U^1, U^2, U^3 $ of the three tori.
\vskip 0.3cm
\noindent{\sl b) The $N_V=N_H=4$ model}

\noindent In order to construct models with lower rank, we start
with the $N=4$ ground state defined above as the $K3\times T^2$
compactification and then  apply
several $Z_2$ freely-acting projections; each one of these
projections
removes half of the $Z_2^{(\rm o)}$-twisted vector multiplets
without changing the Euler number of the compactification manifold,
$\chi\equiv 2(h^{1,1}-h^{2,1})=2(N_V-N_H)=0$, and leads therefore
to models with $N_V=N_H=4$ and 2. It is however important to
stress that this procedure,
already developed in
$N=4$ models \cite{6auth}, can be implemented provided we start from
an orbifold point of the $K3$, namely from the $N=4$
compactification
manifold $\left(T^4/Z^{(\rm o)}_2\right)\times
T^2$. This choice is no longer dictated by convenience.

The model with $N_V=N_H=4$ is constructed by modding out an extra
$Z_2^{D}$ symmetry; the compactification manifold is therefore
${\cal{M}}^{N_V=4}={\cal M}^{N_V=8}/Z_2^{D}$. In order to describe
the action of this projection, we choose special coordinates of the
compact space, in terms of which the
$T^4$ torus is described as a product of circles. There is then a
$(D_4)^4$ symmetry generated by the elements
$D$ and $\tilde{D}$, which act on each $S^1/Z_2$ block as \cite{dvv}:
\be
\eqalign{D:\ \ (\sigma_+,\sigma_-,V_{nm})& \to  \ (\sigma_-,\sigma_+,
(-)^m V_{nm}) \cr
 \tilde{D}:\ \ (\sigma_+,\sigma_-,V_{nm})& \to  (-\sigma_+,\sigma_-,
(-)^n V_{nm}),\cr}
\label{Dop}
\ee
where $\sigma_+$ and $\sigma_-$ are the 2 twist fields of
$S^1/Z_2$, and $V_{nm}$ are the untwisted world-sheet instantons
labelled by the momentum $m$ and the winding $n$ of the
$\Gamma_{1,1}$ lattice. In the
untwisted sector of $T^4/Z^{(\rm o)}_2$, $D$ and $\tilde{D}$ act as
ordinary shifts on the
$\Gamma_{1,1}$ sublattice of the $\Gamma_{4,4}$.

Orbifolding by  $Z^{D}_2$ the original $\left(T^4/Z^{\rm
(o)}_2\right)\times T^2$, we obtain  an
$N=4$ ground state with a reduced number of vector multiplets.
The $Z^{D}_2$ acts on the $T^4/Z^{(\rm o)}_2$  part as a
$D$-operation, while it acts on $T^2$ as an ordinary  shift.
In the $\left(\left(T^4/Z^{\rm
(o)}_2\right)\times T^2\right)\Big/Z_2^{D}$ orbifold, the number of
the $Z^{(\rm o)}_2$-twisted vector multiplets (16) is reduced to 8 by
the extra $Z_2^{D}$ projection, without breaking the $N=4$
supersymmetry further \cite{6auth}. To reduce $N=4$ to $N=2$ we
perform a
further orbifold by using $Z^{(\rm f)}_2$ as in the $N_V=N_H=8$
model above. The $N=2$ ground state obtained in this way is a
compactification on the manifold
${\cal M}^{N_V=4}= \left(\left(T^4/Z^{(\rm o)}_2\right)\times
T^2\right)\Big/\left( Z_2^{(\rm f)}\times Z_2^{D}\right)$ and has
$N_V=N_H=4$. The massless spectrum consists, apart from
the gravity and tensor multiplets, which include a graviphoton and a
dilaton as before, of 7 vector multiplets (4 coming from the twisted
sector) and 7 hypermultiplets (4 twisted).

The partition function based on ${\cal M}^{N_V=4}$ is still given in
Eq. (\ref{zII}), but now with
\be
\Gamma_{6,6}^{N_V=4}\ar{H^{\rm o},H^{\rm f}}{G^{\rm o},G^{\rm
f}}={1\over 2}\sum_{H,G}
\Gamma_{2,2}^{(1)} \ar{H^{\rm f}\vert H}{G^{\rm f}\vert G}
\,
\Gamma_{2,2}^{(2)} \ar{H^{\rm o}\vert H^{\rm f},H}{G^{\rm o}\vert
G^{\rm f},G}
\,
\Gamma_{2,2}^{(3)} \ar{H^{\rm o}+H^{\rm f}\vert H^{\rm f}}{G^{\rm
o}+G^{\rm f}\vert G^{\rm f}}\, ,
\label{z6644}
\ee
where $(H,G)$ correspond to the $Z_2^{D}$-projection.
We introduced the $(2,2)$ conformal blocks
$\Gamma_{2,2} \ar{h\vert h',h''}{g\vert g',g''}$, where, as
previously,
$(h,g)$ refer to
the twist, while $(h',g')$ and $(h'',g'')$  correspond to shifts
along two circles of $T^2$. The non-vanishing components are
$$
\Gamma_{2,2} \ar{h\vert h',h''}{g\vert g',g''}
=
{4\,  \vert\eta \vert^6\over
\left\vert
\vartheta{1+h\atopwithdelims[] 1+g}\,
\vartheta{1-h\atopwithdelims[] 1-g}
\right\vert}
\nn
$$
for $(h',g')=(h'',g'')=(0,0)$
or $(h',g')=(h,g) \ (h'',g'')=(0,0)$
or $(h',g')=(0,0) \ (h'',g'')=(h,g)$
or $(h',g')=(h'',g'')=(h,g)$, and
$$
\Gamma_{2,2} \ar{0\vert h',h''}{0\vert g',g''}
=
\Gamma_{2,2} \ar{h',h''}{g',g''}\, ,
\nn
$$
where $\Gamma_{2,2} \ar{h',h''}{g',g''}$ is the lattice
sum of a $(Z_2 \times Z_2)$-twisted torus.
\vskip 0.3cm
\noindent{\sl c) The $N_V=N_H=2$ model}

\noindent The model with $N_V=N_H=2$ is obtained by
modding out a
further $Z_2^{D}$-symmetry, which acts on another circle of the first
and the third torus:
\be
\Gamma_{6,6}^{N_V=2}\ar{H^{\rm o},H^{\rm f}}{G^{\rm o},G^{\rm f}}=
{1\over 4}\sum_{H,G}\sum_{H',G'}
\Gamma_{2,2}^{(1)} \ar{H^{\rm f}\vert H,H'}{G^{\rm f}\vert G,G'}
\,
\Gamma_{2,2}^{(2)} \ar{H^{\rm o}\vert H^{\rm f},H}{G^{\rm o}\vert
G^{\rm f},G}
\,
\Gamma_{2,2}^{(3)} \ar{H^{\rm o}+H^{\rm f}\vert H^{\rm f},H'}{G^{\rm
o}+G^{\rm f}\vert G^{\rm f},G'}
\label{z66}
\ee
(notice that the two $Z_2^{D}$-projections commute).
The resulting massless spectrum consists of the gravity and tensor
multiplets plus 5 vector multiplets (2 from the  twisted sector), and
5 hypermultiplets (2 from the twisted sector).

In all the above models, the
realization of the $N=2$ supersymmetry plays a key
role in the search of heterotic duals, as we will see in Section
\ref{het}. Indeed, by using  the techniques developed in Refs.
\cite{kk, solving}, we can indeed show that these models actually
possess
a spontaneously broken $N=4$ supersymmetry through a
Higgs phenomenon, due to the free action of $Z_2^{(\rm f)}$.
The restoration of the 16 supersymmetric charges
(2 extra massless gravitinos) takes place
in some appropriate limits of the moduli according to the precise
shifts in the lattices. It is accompanied by a logarithmic instead of
a linear blow-up of various thresholds \cite{kk,solving,kou,kkprn},
which is nothing
but an infrared artifact due to an accumulation of massless states.
These can be lifted by introducing an infra-red cut-off $\mu$
larger than the two
gravitino masses. The thresholds thus vanish as expected in the
limit in which supersymmetry
is extended to $N=4$ as ${m_{3/2}/ \mu} \to 0$ .

\subsection{Helicity supertraces and the $R^2$ corrections}

Usually string ground states
are best described by writing their
(four-dimensional) helicity-generating partition functions. Moreover,
since our motivation is eventually to compute couplings associated
with interactions such as $R^2$,  we need in general to evaluate
amplitudes
involving operators such as $i \left( X^{3}
\buildrel{\leftrightarrow}\over{\partial}
X^{4} + 2 \psi^{3}\psi^{4}\right)\bJ^k$  \cite{infra},
where $\bJ^k$ is an appropriate right-moving
current and the left-moving factor corresponds to the left-helicity
operator (the right-helicity operator is the antiholomorphic
counterpart of the latter). We will not expand here on the various
operators of this kind, or on the
procedures that have been used in order to
calculate their correlation functions exactly (i.e. to all orders in
$\alpha '$  and without infrared ambiguities \cite{infra});
details will be given in Section \ref{hetthr}, when analysing some
specific
$R^2$ corrections for the heterotic duals of the models under
consideration. Here we will restrict ourselves to the
helicity-generating partition functions since these allow for a
direct computation of perturbative type II $R^2$ corrections. They
are defined as:
\be
Z(v,\bar v)=\Trp
q^{L_0-{c\over 24}}\,
\bar q^{\bar L_0 - {\bar c \over 24}}\,
e^{2\pi i \left(v Q-\bar v \bQ \right)}
\, ,
\label{hel}
\ee
where the prime over the trace excludes the zero-modes related to the
space-time coordinates (consequently
$Z(v,\bar v)\vert_{v=\bar v=0}= \t_2 Z$, where $Z$ is the vacuum
amplitude) and $Q,\bQ$
stand for the left- and right-helicity contributions to
the four-dimensional physical helicity. Various helicity
supertraces are finally obtained by taking appropriate
derivatives of (\ref{hel}):
\be
B_{2n}=\left\langle \left( Q+\bQ
\right)^{2n}\right\rangle_{\rm genus-one}
\ee
is obtained by acting
on $Z(v,\bar v)$ with
$\frac{1}{(-4\pi^2)^n}\left(\partial_v-\partial_{\bar v}\right)^{2n}$
at $v = \bar v = 0$.

For the models at hand (see Eq. (\ref{zII})), after some algebra,
(\ref{hel}) reads:
\be
Z_{\rm II}^{N_V}(v,\bar v)=
{\xi(v)\, \bar\xi(\bar v)\over \vert\eta\vert^{4}}
{1\over 4}\, \sum_{H^{\rm o},G^{\rm o}}\sum_{H^{\rm f},G^{\rm f}}
{\Gamma_{6,6}^{N_V}\ar{H^{\rm o},H^{\rm f}}{G^{\rm o},G^{\rm f}}
\over \vert\eta\vert^{12}}\,
Z_{\rm L}^{\rm F}\ar{H^{\rm o},H^{\rm f}}{G^{\rm o},G^{\rm f}}(v)\,
Z_{\rm R}^{\rm F}\ar{H^{\rm o},H^{\rm f}}{G^{\rm o},G^{\rm f}}(\bar
v)\, ,
\label{hII}
\ee
where
$Z_{\rm L}^{\rm F}\ar{H^{\rm o},H^{\rm f}}{G^{\rm o},G^{\rm f}}(v)$
and
$Z_{\rm R}^{\rm F}\ar{H^{\rm o},H^{\rm f}}{G^{\rm o},G^{\rm f}}(\bar
v)$
denote the contribution of the 8 left- and 8 right-moving
world-sheet
fermions $\psi^{\mu}, \Psi^I$ and $\bar{\psi}^{\mu}, \bar{\Psi}^I$;
$\Psi^I$ and $\bar{\Psi}^I$ are the 6 left- and 6 right-moving
fermionic degrees of freedom of the  six-dimensional internal space.
The arguments $v$ and $\bar v$ are due to $\psi^{\mu}$ and  ${\bar
\psi}^{\mu}$.
By using the Riemann
identity of theta functions, one can perform the summation over the
spin structures with the result:
\be
Z_{\rm L}^{\rm F}\ar{H^{\rm o},H^{\rm f}}{G^{\rm o},G^{\rm f}}(v)=
{1\over \eta^4} \,
\vartheta \ar{1}{1}\left({v \over 2}\right)\,
\vartheta \ar{1-H^{\rm o}}{1-G^{\rm o}}\left({v \over 2}\right)\,
\vartheta \ar{1-H^{\rm f}}{1-G^{\rm f}}\left({v \over 2}\right)\,
\vartheta \ar{1+H^{\rm o}+H^{\rm f}}{1+G^{\rm o}+G^{\rm f}}\left({v
\over 2}\right)
\label{ZFL}
\ee
and
\be
Z_{\rm R}^{\rm F}\ar{H^{\rm o},H^{\rm f}}{G^{\rm o},G^{\rm f}}(\bar
v)=
{1\over \bar \eta^4} \,
\bar \vartheta \ar{1}{1}\left({\bar v \over 2}\right)\,
\bar \vartheta \ar{1-H^{\rm o}}{1-G^{\rm o}}\left({\bar v \over
2}\right)\,
\bar \vartheta \ar{1-H^{\rm f}}{1-G^{\rm f}}\left({\bar v \over
2}\right)\,
\bar \vartheta \ar{1+H^{\rm o}+H^{\rm f}}{1+G^{\rm o}+G^{\rm
f}}\left({\bar v
\over 2}\right)\, .
\label{ZFR}
\ee
Finally
$$
\xi(v)=\prod_{n=1}^{\infty}{(1-q^n)^2\over \left(1-q^n e^{2\pi
iv}\right)\left(1-q^n e^{-2\pi iv}\right)}=
{\sin\pi v\over \pi}{\vartheta_1'(0)\over
\vartheta_1(v)}
$$
counts the helicity contributions of the space-time bosonic
oscillators.

A straightforward computation based on the techniques developed
so far shows that
$B_2 = 0$ for all the models under consideration, as expected.
Indeed, in all the $N=2$ type II $Z_2 \times Z_2$ symmetric
orbifolds,
$B_2$ can receive a non-zero contribution only from the
$N=(1,1)$ sectors of the orbifold ( see Ref. \cite{gkpr}). The
internal coordinates in these sectors are twisted; all corrections
are
therefore moduli-independent and are coming from the massless states
only. One finds $B_2=B_2\vert_{\rm massless}= N_V-N_H$, which
vanishes
in all models we are considering here.

On the
other hand, $B_4$  receives non-zero contributions from the $N=(2,2)$
sectors of the orbifold. We find \footnote{The prime summation
over $(h,g)$ stands for $(h,g) =
\{(0,1),(1,0),(1,1)\} $.}:
\be
B_4^{N_V=8}= 18\, \Gamma_{2,2}^{(1)}+
6\sum_{i=2,3} \sump \Gamma_{2,2}^{(i)}\ar{0\vert h}{0\vert g}
\, ,
\label{b488}
\ee
\ba
B_4^{N_V=4}&=& 9\, \Gamma_{2,2}^{(1)}+
3 \sump \Gamma_{2,2}^{(1)}\ar{0\vert h}{0\vert g}
+
3 \sump \Gamma_{2,2}^{(2)}\ar{0\vert h,0}{0\vert g,0}
\nn \\
&&+3\sump\left(\Gamma_{2,2}^{(2)}\ar{0\vert h,h}{0\vert g,g}
+2\,  \Gamma_{2,2}^{(3)}\ar{0\vert h}{0\vert g} \right)
\, ,
\label{b444}
\ea
\ba
B_4^{N_V=2}&=& \frac{9}{2} \,\Gamma_{2,2}^{(1)}+
\frac{3}{2}\sump \left(\Gamma_{2,2}^{(1)}\ar{0\vert h,0}{0\vert g,0}
+\Gamma_{2,2}^{(1)}\ar{0\vert 0,h}{0\vert 0,g}
+\Gamma_{2,2}^{(1)}\ar{0\vert h,h}{0\vert g,g}
\right)\nonumber\\
&&+3\sum_{i=2,3}\sump\left(\Gamma_{2,2}^{(i)}
\ar{0\vert h,0}{0\vert g,0}
+\Gamma_{2,2}^{(i)}\ar{0\vert h,h}{0\vert g,g} \right)
\, .
\label{b422}
\ea
The massless contributions are in agreement with the generic
result of the $N=2$ supergravity:
\be
\left. B_4\right\vert_{\rm massless}=18 + \frac{7 N_V-N_H}{4}\, .
\label{B4m}
\ee

We now turn to the actual computation of $R^2$ corrections. The
four-derivative gravitational corrections we will consider here are
precisely those that were analysed in the framework of $N=4$ ground
states of Ref. \cite{6auth}. On the type II side,
there is no tree-level contribution to these operators, and
the $R^2$
corrections are related to the insertion of the two-dimensional
operator $2Q^2 \bQ^2$ in the one-loop partition
function. In the models at hand, where supersymmetry is
realized symmetrically and where, moreover, $N_V=N_H$, the
contribution of $N=(1,1)$ sectors to $B_4$ vanishes, and therefore
$\left\langle 2 Q^2 \bQ^2\right\rangle$ can be
identified with $B_4 / 3$. The massless contributions of the latter
give rise to an infrared logarithmic behaviour $2b_{\rm
II}\log\left( M^{(\rm II)\,2}  \Big/ \mu^{(\rm II)\,2}\right)$
\cite{infra,delgrav}, where
$M^{(\rm II)}\equiv\frac{1}{\sqrt{\alpha'_{\rm II}}}$ is
the type II string scale and $\mu^{(\rm II)}$ is the infrared cut-off.
Besides this running, the one-loop correction contains, as usual, the
thresholds $\Delta_{\rm II}$, which account for the infinite tower of
string modes.

The one-loop corrections of the $R^2$ term are related to the
infrared-regularized
genus-one
integral of $ B_4 / 3$. There is however a subtlety: in the type IIA
string, these $R^2$ corrections depend on the K\"ahler moduli
(spanning the vector manifold), and are independent of the
complex-structure moduli (spanning the scalar manifold):
\be
\partial_{T^i}\Delta_{\rm IIA}=\frac{1}{3}\ifd
\partial_{T^i} B_4
\ , \ \
\partial_{U^i}\Delta_{\rm IIA}=0\, .
\label{IIAthr}
\ee
In  the  type IIB string, the roles of $T^i$ and $U^i$ are
interchanged:
\be
\partial_{U^i}\Delta_{\rm IIB}=\frac{1}{3}\ifd
\partial_{U^i} B_4
\ , \ \
\partial_{T^i}\Delta_{\rm IIB}=0\, .
\label{IIBthr}
\ee
The above properties of $\Delta_{\rm IIA,B}$  generalize the results
of Ref. \cite{6auth}, which are valid for the $R^2$ corrections in
$N=4$ models with reduced-rank gauge group. They also account for the
massless contribution
$b_{\rm II}\log\left( M^{(\rm II)\,2}  \Big/ \mu^{(\rm II)\,2}\right)$ 
being half the infrared-regularized massless contribution of the
integral of
$B_4 / 3$ over the fundamental domain.

In order to perform the integration, we have to specify what the
translations induced by $Z_2^{(\rm f)}$, $Z_2^{D}$ and $Z_2^{D'}$
are. For definiteness we will choose the same half-unit shifts on
momenta
for all three complex planes: the first shift ($Z_2^{(\rm f)}$) will
act on the first momentum (insertion
of $(-1)^{m_1^i}$ in the $i$th plane);
the second shift ($Z_2^{D}$ or $Z_2^{D'}$,
when necessary, as in models $(4,4)$ and $(2,2)$) will act on the
second momentum (insertion of $(-1)^{m_2^i}$).
Given this choice, the final result for the running now reads:
\ba
{16\, \pi^2\over g^2_{\rm grav}(\mu^{(\rm II)})} &=& - 
\frac{3N_V}{4}
\log {\mu^{(\rm II)} \Im T^1\over M^{(\rm II)}}\,\left\vert
\eta\left(T^1\right)
\right \vert^4
-\left(2-{N_V \over 4}\right)\log { \mu^{(\rm II)}\Im T^1\over
M^{(\rm II)}}\,
\left\vert \vartheta_4\left(T^1\right)
\right \vert^4\nonumber \\
&&-2\log {\mu^{(\rm II)}\Im T^2\over M^{(\rm II)}}\,\left\vert
\vartheta_4\left(T^2\right)
\right \vert^4 - 2\log {\mu^{(\rm II)}\Im T^3\over
M^{(\rm II)}}\,\left\vert
\vartheta_4\left(T^3\right)
\right\vert^4 + \ {\rm const.} \,  \label{thrint}
\ea
Expression (\ref{thrint}) deserves several comments:

(\romannumeral1) The shifts  on the $\Gamma^{(i)}_{2,2}$
lattices break the $SL(2,Z)_{T^i}$ duality groups. The residual
subgroup depends in fact on the kind of shifts performed
(see Refs. \cite{ 6auth, solving,kkprn}).

(\romannumeral2) The $SL(2,Z)_{T^1}$ remains
unbroken only in the model with $N_V=N_H=8$;
in the other situations all $SL(2,Z)_{T^i}$ are necessarily
broken for all $i=1,2,3$.

(\romannumeral3) The $N=4$ restoration limit corresponds to
$T^2,T^3\to \infty$. For the specific choice of translations
we are considering, the mass of the 2 extra gravitinos is given by:
\be
m^2_{3/2}={ 1 \over 4 \Im T^2 \Im U^2 }+
{ 1 \over 4 \Im T^3 \Im U^3 }   \ , \ \
{\rm for\ } N_V=8 \, ,
\label{mg8}
\ee
\be
m^2_{3/2} = { 1 \over 4 \Im T^2 \Im U^2 }+{ \Im U^2 \over 4 \Im
T^2  }
+ { 1 \over 4 \Im T^3 \Im U^3 }   \ , \ \
{\rm for\ } N_V=4
\label{mg4}
\ee
and
\be
m^2_{3/2} = { 1 \over 4 \Im T^2 \Im U^2 }+{ \Im U^2 \over 4 \Im
T^2  }
+ { 1 \over 4 \Im T^3 \Im U^3 }+ { \Im U^3 \over 4 \Im T^3 } \ ,
\ \ {\rm for\ } N_V=2\, .
\label{mg2}
\ee
Owing to the effective restoration of $N=4$ supersymmetry in this
limit, there is no  linear behaviour either in $\Im T^2$ or in  $\Im
T^3$; the
remaining contribution is  logarithmic:
\be
{16 \, \pi^2 \over g_{\rm grav}^2(\mu^{(\rm II)})} \ \underarrow{\Im T^2,
\Im T^3
\to \infty} \
-2 \log \Im T^2 - 2 \log \Im T^3\, .
\label{limn4}
\ee

(\romannumeral4) The threshold corrections
diverge linearly in the large $\Im T^1$ limit.
\be
{16 \, \pi^2 \over g^2_{\rm  grav} \left(\mu^{(\rm II)} \right) }
 \limit{\longrightarrow}{\Im T^1 \to \infty} 
 { \pi N_V \over 4} \Im T^1
-\left({2}+{N_V\over 2}\right) \log \Im T^1
+ \left( 6 + {N_V \over 2} \right)\,
\log {M^{(\rm II)} \over \mu^{(\rm II)}}\, .
\label{limn2}
\ee
Observe that the coefficients of the linear and  logarithmic term have
a different dependence on $N_V$. 
The coefficient
of the third term, $6+{N_V \over 2}$, is actually
$B_4 / 3$,
while that of the second term, $2+{N_V \over  2}$, is
$\left( B_4 -6b_{\rm grav}\right)/3$,
where by $B_4$ we actually intend the massless
contribution to this
quantity:
$B_4\vert_{\rm massless}$ given in Eq. (\ref{B4m});
$b_{\rm grav}$ is the ``gravitational beta-function'',
$b_{\rm grav}=\left( 24-N_V+N_H \right)/ 12$, 
as computed in field theory.
Notice that in general, when $N_V \neq N_H$, the relevant quantities that 
appear in the above corrections are:
\be
{B_4-B_2 \over 3} -2b_{\rm grav}=
2+{5N_V+N_H \over 12}~~~~~ {\rm and}~~~~~ {B_4-B_2 \over 3}
= 6+{N_V+N_H \over 4},
\label{run}
\ee
where by $B_4$ and $B_2$ we intend, as before, 
the massless contribution to these quantities
($B_2\vert_{\rm massless}=N_V-N_H$).

(\romannumeral5) Under $T^i \leftrightarrow U^i$ interchange, we
obtain the results of a mirror type IIB model.

\section{Heterotic duals}\label{het}

\subsection{Outline}

Our scope is now to determine  the heterotic duals
of the  type II ground states with $N_V=8,4,2$ discussed
in the previous section. In view of this construction, some
basic properties and  requirements have to be settled,
namely:

(\romannumeral1) The  $N=2$ heterotic models must have the
same massless spectrum as their type II duals.

(\romannumeral2) The type II dual of the heterotic dilaton
$S_{\rm Het}$ is one of the type IIA(B) moduli, $T_D$, and belongs to
a vector multiplet. Thus, the perturbative heterotic
limit  ($S_{\rm Het}$ large) corresponds in type II to the limit of
large
$T_D$. This implies the identification of $T_D$ with  $T^1$ in
our type IIA constructions.

(\romannumeral3)  The $N=2$ heterotic ground state must describe a
spontaneously broken phase of an $N=4$ model. The limit in which  the
$N=4$ is restored
corresponds to large perturbative type IIA vector-multiplet  moduli
$T^2, T^3$
(see Eqs. (\ref{mg8})--(\ref{mg2}) and (\ref{limn4})). Therefore,
we must identify the  $T^2$ and $ T^3 $ of type IIA  with two
perturbative
moduli of the heterotic duals, $T_{\rm Het}$ and $U_{\rm Het}$.

(\romannumeral4)  On the type II side, the $T$-duality group is a
{\it subgroup} of
\be
SL(2,Z)_{T^1} \times SL(2,Z)_{T^2} \times SL(2,Z)_{T^3}\, .
\ee
Moreover, the duality symmetries of $T^1\equiv S_{\rm Het} $
translate into the  non-perturbative instanton-duality properties
on the heterotic side, while the symmetries of the  moduli
$T^2\equiv T_{\rm Het}$ and
$T^3\equiv U_{\rm Het}$ appear in the heterotic duals as perturbative
$T$-duality symmetries.
These properties can be summarized as follows:
\be
\begin{array}{lcc}
$type IIA:$ & SL(2,Z)_{T^1} \ , & SL(2,Z)_{T^2} \times
SL(2,Z)_{T^3}\\
            & \updownarrow    &    \updownarrow   \\
$heterotic:$ & SL(2,Z)_{S} \ , & SL(2,Z)_{T} \times SL(2,Z)_{U}\, ,
\end{array}
\label{st}
\ee
where the $SL(2,Z)$'s can be broken to some $\Gamma(2)$ subgroups.

On the heterotic side, the general expression for the
helicity-generating function of $Z_2$-orbifold models is
\be
Z_{\rm Het}^{N_V}(v,\bar v)=
{\xi(v)\, \bar\xi(\bar v)\over \vert\eta\vert^{4}}\,
{1\over 2}\, \sum_{H^{\rm f},G^{\rm f}}
Z_{6,22}^{N_V}\ar{H^{\rm f}}{G^{\rm f}}\,
Z_{\rm L}^{\rm F}\ar{H^{\rm f}}{G^{\rm f}}(v)\, ,
\label{hH}
\ee
where $Z_{\rm L}^{\rm F}\ar{H^{\rm f}}{G^{\rm f}}(v)$ is
the contribution of the 8 left-moving
world-sheet fermions $\psi^{\mu},\Psi^I$ in the light-cone gauge
(see (\ref{ZFL}) with $H^{\rm o}=G^{\rm o}=0$) and
$Z_{6,22}^{N_V}\ar{H^{\rm f}}{G^{\rm f}}$ accounts for the (6,22)
compactified coordinates. In order to allow for a comparison with the
type II orbifolds, the
$\Gamma_{2,2}(T,U)$ shifted lattice needs to be
separated. It is also necessary to choose
special values of the remaining Wilson-line moduli,
$Y_I, I=1,\ldots , N_V$ of the $\Gamma_{2,2+N_V}$ lattice,
which break the
gauge group to $U(1)$ factors. At such points,
$Z_{6,22}^{N_V}\ar{H^{\rm f}}{G^{\rm f}}$ takes the
following form:
\be
Z_{6,22}^{N_V}\ar{H^{\rm f}}{G^{\rm f}}=
{1\over 2^{n+1}}\, \sum_{\vec h, \vec g}
{\Gamma_{2,2} \ar{H^{\rm f},\vec h}{G^{\rm f},\vec g}\over
\vert\eta\vert^{4}}\,
{\Gamma_{4,4} \ar{H^{\rm f}\vert \vec h}{G^{\rm f}\vert \vec g}\over
\vert\eta\vert^{8}}\,\sum_{\g, \d}\,
{\overline \Phi}^V \ar{H^{\rm f},\g, \vec h}{G^{\rm f},\d, \vec g}\,
{\overline \Phi}^H \ar{H^{\rm f},\g, \vec h}{G^{\rm f},\d, \vec g}\,
,\label{hH622}
\ee
where $(\vec h, \vec g)$ denote  either the values of the Wilson
lines or the heterotic duals of the $Z^{D}_2$ operations; $n$
is the number of projections (dimension of the vectors $(\vec h,
\vec g)$) needed to reach the correct $N_V$. The specific lattice
shifts define the modular-transformation properties of the
multi-shifted two-torus lattice sum
$\Gamma_{2,2}\ar{H^{\rm f},\vec h}{G^{\rm f},\vec g}$, and
must fit with the transformation properties of the $\bPhi$'s
in order to lead to the proper modular-covariance properties of
$Z_{6,22}^{N_V}\ar{H^{\rm f}}{G^{\rm f}}$.

Taking into account the above considerations, we will now construct
the heterotic duals of the various reduced-rank type II models.

\subsection{Construction of the heterotic reduced-rank models}

\noindent {\sl a) The $N_V=N_H=8$ model}

\noindent Our starting point is the  $N=4$ heterotic
ground state obtained by compactification on $T^4 \times T^2$
and its type II dual on $K3 \times T^2$.
In order to reduce the $N=4$ supersymmetry
to $N=2$ we must define an appropriate   $Z_2^{(\rm f)}$
freely-acting orbifold, which simultaneously reduces
by a factor of 2 the rank of the gauge group.

To properly define the $Z_2^{(\rm f)}$ action, we work
at a point of the moduli space where the $E_8\times E_8$ gauge
group is broken to
$\left( SU(2)_{k=1}^{(1)}\times
SU(2)_{k=1}^{(2)}\right)^8\simeq SO(4)^8_{k=1}$.
This point is reached by
switching on appropriate Wilson lines of the $(6,22)$ lattice.
The action of $Z_2^{(\rm f)}$ then amounts to
(\romannumeral1) a translation on $T^2$, which produces a
half-unit-vector shift in the $(2,2)$ lattice;
(\romannumeral2) a $Z_2^{(\rm f)}$ symmetric twist on $T^4$, which
breaks the $N=4$ supersymmetry to $N=2$;
(\romannumeral3) a  pair-wise interchange of $SU(2)^{(1)}$ with
$SU(2)^{(2)}$.

We can give an explicit expression for the partition function
of this orbifold by using the following $SO(4)$ twisted  characters:
\be
F_{1}\ar{\g,h}{\d, g} \equiv {1\over {\eta}^2}\,
{\vartheta}^{1/2}
\ar{\g+h_1}{\d+g_1}\,
{\vartheta}^{1/2}
\ar{\g+h_2}{\d+g_2}\,
{\vartheta}^{1/2}
\ar{\g+h_3}{\d+g_3}\,
{\vartheta}^{1/2}
\ar{\g-h_1-h_2-h_3}{\d-g_1-g_2-g_3}
\ee
and
\be
F_{2}\ar{\g, h}{\d, g} \equiv {1\over {\eta}^2}\,
{\vartheta}^{1/2}
\ar{\g}{\d}\,
{\vartheta}^{1/2}
\ar{\g+h_1-h_2}{\d+g_1-g_2}\,
{\vartheta}^{1/2}
\ar{\g+h_2-h_3}{\d+g_2-g_3}\,
{\vartheta}^{1/2}
\ar{\g+h_3-h_1}{\d+g_3-g_1}\, ,
\ee
where we introduced the notation (valid all over the paper)
$h\equiv (h_1,h_2,h_3)$ and similarly for $g$.
Under $\tau\to \tau+1$, $F_{I}$
transform as:
\ba
F_{1}\ar{\g, h}{\d, g} & \to &
F_{1}\ar{\g, h}{\g+\d +1, h + g} \nn \\
&& \times
\exp -{i\pi\over 4}
\left({2\over 3} - 4\g +2 \gamma^2+h_1^2+h_2^2 +h_3^2 + h_1 h_2+h_2
h_3+h_3 h_1 \right)
\\
F_{2}\ar{\g, h}{\d, g} & \to &
F_{2}\ar{\g, h}{\g+\d +1, h + g} \nn \\
&& \times
\exp -{i\pi\over 4}
\left({2\over 3} - 4\g +2 \gamma^2+h_1^2+h_2^2 +h_3^2 - h_1 h_2-h_2
h_3-h_3 h_1 \right)\, .
\ea
Notice that $\bF_{I}$ are $c=(0,2)$ conformal characters of 4
different right-moving Isings. In the fermionic language \cite{abk}
this is a
system
of 4 right-moving real fermions with different boundary
conditions.
All currents $\bar{J}^{IJ}={\bar \Psi}^I{\bar \Psi}^J$ are twisted
and therefore
the initial $SO(4)$ is broken. On the other hand, when
$\bF_{I}$ is raised to a power $n$, the gauge group becomes
$SO(n)^4$:
\be
\bF_{I}^n\ar{\g, h}{\d, g} \to SO(n)^4\, .
\ee
In this framework, the partition function of the model with
$N_V=N_H=8$ is given by (\ref{hH}) and (\ref{hH622}), with
$\Phi^{V}$ and $\Phi^H$ expressed in terms of combinations of
$F_{1}\ar{\g, h}{\d, g}$ and $F_{2}\ar{\g, h}{\d, g}$:
\be
\Phi^V\ar{\g,\vec h}{\d,\vec g}
= F_{1}^2\ar{\g, h}{\d, g}\, F_{2}^2\ar{\g, h}{\d, g}
\label{88fiv}
\ee
(no dependence on $(H^{\rm f},G^{\rm f})$),
which leads to a group \footnote{Observe that the right moving gauge
group is systematically $U(1)^2 \times G$.}
$G=U(1)^8$ and therefore
$N_V=8$, whereas
\be
\Phi^H\ar{H^{\rm f},\g,\vec  h}{G^{\rm f},\d,\vec g}
= F_{1}\ar{\g', h}{\d', g}\,
F_{1}\ar{\g'+H^{\rm f}, h}{\d'+G^{\rm f}, g}\,
F_{2}\ar{\g', h}{\d', g}\,
F_{2}\ar{\g'+H^{\rm f}, h}{\d'+G^{\rm f}, g}\, ,
\label{88fih}
\ee
leading to $N_H=8$  (here
$\vec h \equiv (h_1, h_2, h_3,h_4)=(h,h_4)$, $\vec g \equiv (g_1, g_2,
g_3,g_4)=(g,g_4)$
and $(\g',\d')=(\g+h_4,\d+g_4)$).
For this model, we must use the simply-shifted $(2,2)$ lattice sum
$\Gamma_{2,2} \ar{H^{\rm f}}{G^{\rm f}}$, where
the shift is
asymmetric
on one circle $S^1$, projection $(-)^{(m_2+n_2) G^{\rm f}}$, in order
to cancel
the phase of $Z_{\rm L}^{\rm F}\ar{H^{\rm f}}{G^{\rm f}}$ under
modular transformations (this projection was referred to as ``XIII"
in
Ref. \cite{kkprn}, where the various lattice shifts were discussed in
detail). On the other hand, the
$h_i$ shifts in $\Gamma_{4,4}$ must be symmetric:
$(-)^{M_ig_i}$.

An alternative construction is obtained by shifting with $(H^{\rm
f},G^{\rm f})$ two $U(1)$ factors in $G$:
\ba
\tilde{\Phi}^V \ar{H^{\rm f},\g, \vec h}{G^{\rm f},\, \d, \vec g}
&=&
 {1\over {\eta}^8}\,
{\vartheta}
\ar{\g + h_1 + H^{\rm f}}
   {\d + g_1 + G^{\rm f}}
\,
{\vartheta}
\ar{\g + h_2}
   {\d + g_2}
\,
{\vartheta}
\ar{\g + h_3}
   {\d + g_3}
\,{\vartheta}
\ar{\g - h_1 - h_2 - h_3}
   {\d - g_1 - g_2 - g_3}
 \nonumber \\
&& \times
{\vartheta}
\ar{\g - H^{\rm f}}
   {\d - G^{\rm f}}
\,
{\vartheta}
\ar{\g + h_1 - h_2}
   {\d + g_1 - g_2}
\,
{\vartheta}
\ar{\g + h_2 - h_3}
   {\d + g_2 - g_3}
\,
{\vartheta}
\ar{\g + h_3 - h_1}
   {\d + g_3 - g_1}\, .
\label{88fivt}
\ea
The $(2,2)$ lattice is now double-shifted, and the lattice sum
$\Gamma_{2,2}\ar{H^{\rm f}, h_1}{G^{\rm f}, g_1}$ is
given with the insertion
$(-)^{m_2 G^{\rm f} + n_2 g_1 }$.
Both models, with  $\Phi^V$ and $\tilde{\Phi}^V$, have the same
$N=4$ sector (defined by $(H^{\rm f},G^{\rm f})=(0,0)$, whose
contribution is half the partition function of an $N=4$ model
in which the gauge group $E_8 \times E_8$
is broken to $U(1)^{16}$. In the  $N=2$ sectors,
$(H^{\rm f},G^{\rm f})\neq (0,0)$ and $(h_i, g_i)$ are either
$(0,0)$ or $(H^{\rm f},G^{\rm f})$. This constraint comes from the
$(H^{\rm f},G^{\rm f})$-twisted
sector of $\Gamma_{4,4} \ar{H^{\rm f}\vert \vec h}{G^{\rm f}\vert
\vec
g}$.

As a consistency check, we can now proceed to the computation
of the helicity supertrace
$B_2$.  The ($N=4$)-sector contribution to this quantity vanishes.
For
the model constructed with $\Phi^V$ (Eq. (\ref{88fiv})) we find:
\be
B_2 \left(\Phi^V \right)={1\over \bar{\eta}^{24}}\,
\sumpf
\Gamma_{2,2}^{\lambda=1} \ar{H^{\rm f}}{G^{\rm f}} \,
\bOmega \ar{H^{\rm f}}{G^{\rm f}}\, ,
\label{B288}
\ee
where $\Omega \ar{H^{\rm f}}{G^{\rm f}}$ are the following
analytic functions \footnote{The parameter $\l$, which takes the
values 0 or 1, determines the phases appearing in the modular
transformations of the shifted lattice sums. Under modular
transformations, the functions $\Omega$
acquire phases that are complementary to those coming from the
lattice $\Gamma_{2,2}^{\lambda=1} \ar{H^{\rm f}}{G^{\rm f}}$ that
appears in (\ref{B288}).
This ensures that the spin connection is correctly embedded.
Similarly the $\Omega^{(0)}$'s and $\Omega^{(1)}$'s in (\ref{B288t})
are respectively of type $\l =0 $ and $\l =1 $.}:
\ba
\Omega \ar{0}{1}&=&\hphantom{-}{1\over 16}
\left({\vartheta}_3^8 + {\vartheta}_4^8
+14\, {\vartheta}_3^4\, {\vartheta}_4^4 \right)
{\vartheta}_3^6\,  {\vartheta}_4^6\nn \\
\Omega \ar{0}{1}&=&-{1\over 16}
\left({\vartheta}_2^8 + {\vartheta}_3^8
+14\, {\vartheta}_2^4\, {\vartheta}_3^4 \right)
{\vartheta}_2^6\,  {\vartheta}_3^6\label{Om88} \\
\Omega \ar{0}{1}&=&\hphantom{-}{1\over 16}
\left({\vartheta}_2^8 + {\vartheta}_4^8
-14\, {\vartheta}_2^4\, {\vartheta}_4^4 \right)
{\vartheta}_2^6\,  {\vartheta}_4 ^6\, ;\nn
\ea
the lattice sum
$\Gamma_{2,2}^{\lambda=1} \ar{H^{\rm f}}{G^{\rm f}}$
corresponds to the projection $(-)^{(m_2+n_2) G^{\rm f}}$
(lattice ``XIII").

For the model constructed with $\tilde{\Phi}^V$
(Eq. (\ref{88fivt})) we find instead:
\be
B_2 \left(\tilde{\Phi}^V \right)=
{1\over \bar{\eta}^{24}}\,
\sumpf{1\over 2}\left(
\Gamma_{2,2}^{\lambda=0} \ar{H^{\rm f}}{G^{\rm f}} \,
\bOmega^{(0)} \ar{H^{\rm f}}{G^{\rm f}} +
\Gamma_{2,2}^{\lambda=1} \ar{H^{\rm f}}{G^{\rm f}} \,
\bOmega^{(1)} \ar{H^{\rm f}}{G^{\rm f}}
\right) \, ,
\label{B288t}
\ee
where in this case
\ba
\Omega^{(0)}\ar{0}{1}&=&{\hphantom{-}}
\frac{1}{2}
\left(\th_3^4+\th_4^4\right)
\th_3^8\, \th_4^8
\nonumber\\
\Omega^{(0)}\ar{1}{0}&=&-
\frac{1}{2}
\left(\th_2^4+\th_3^4\right)
\th_2^8\, \th_3^8
\label{Om88a}\\
\Omega^{(0)}\ar{1}{1}&=&{\hphantom{-}}
\frac{1}{2}
\left(\th_2^4-\th_4^4\right)
\th_2^8\, \th_4^8\nonumber
\ea
and
\ba
\Omega^{(1)}\ar{0}{1}&=&{\hphantom{-}}
\th_3^{10}\, \th_4^{10}
\nonumber\\
\Omega^{(1)}\ar{1}{0}&=&-
\th_2^{10}\, \th_3^{10}
\label{Om88b}\\
\Omega^{(1)}\ar{1}{1}&=&-
\th_2^{10}\, \th_4^{10}\, .
\nonumber
\ea
The lattice sums $\Gamma_{2,2}^{\lambda=0} \ar{H^{\rm f}}{G^{\rm f}}$
and
$\Gamma_{2,2}^{\lambda=1} \ar{H^{\rm f}}{G^{\rm f}}$
correspond to simply-shifted lattices with projections respectively
$(-)^{m_2 G^{\rm f}}$ and $(-)^{(m_2+n_2) G^{\rm f}}$
(in \cite{kkprn} they are referred to respectively as ``II'' and
``XIII'').

It is easy to check that, in both constructions, the massless
contribution to the $B_2$ vanishes, as it should for models where
$N_V = N_H$.
\vskip 0.3cm
\noindent{\sl b) The $N_V=N_H=4$  heterotic model}

\noindent This model is realized by using the following functions:
\be
\Phi^V\ar{\g,\vec h}{\d,\vec g}={1\over \eta^4}\,
\vartheta \ar{\g}{\d}\,
\vartheta \ar{\g+h_1-h_2}{\d+g_1-g_2}\,
\vartheta \ar{\g+h_2-h_3}{\d+g_2-g_3}\,
\vartheta \ar{\g+h_3-h_1}{\d+g_3-g_1}
\label{44fiv}
\ee
(no dependence on $(H^{\rm f},G^{\rm f})$) and
\ba
\Phi^H\ar{H^{\rm f},\g,\vec h}{G^{\rm f},\d,\vec g}&=&
F_1\ar{\g+h_4+h_5,h}{\d+g_4+g_5,g}\,
F_1\ar{\g+h_4+h_5+H^{\rm f},h}{\d+g_4+g_5+G^{\rm f},g}\nn \\
&&\times 
F_1\ar{\g+h_5,h}{\d+g_5,g}\,
F_1\ar{\g+h_4+H^{\rm f},h}{\d+g_4+G^{\rm f},g}\nn \\
&&\times 
F_2\ar{\g+h_5,h}{\d+g_5,g}\,
F_2\ar{\g+h_4,h}{\d+g_4,g}\, .
\label{44fih}
\ea
Now $\vec h \equiv (h_1,h_2,h_3,h_4,h_5)=(h,h_4,h_5)$ and similarly
for $\vec g$.
The shift in the $(2,2)$ lattice is asymmetric and along a single
circle with projection
$(-)^{(m_2+n_2)G^{\rm f}}$, while it is symmetric
in the $(4,4)$ block: $(-)^{M_ig_i}$,  $i=1,\ldots,4$.

Again an alternative construction exists for $\Phi^V$; it is given by
\be
{\tilde \Phi}^V\ar{H^{\rm f},\g,\vec h}{G^{\rm f},\, \d,\vec g}=
{1\over \eta^4}\,
\vartheta \ar{\g-H^{\rm f}}{\d-G^{\rm f}}\,
\vartheta \ar{\g+h_1-h_2+H^{\rm f}}{\d+g_1-g_2+G^{\rm f}}\,
\vartheta \ar{\g+h_2-h_3}{\d+g_2-g_3}\,
\vartheta \ar{\g+h_3-h_1}{\d+g_3-g_1}\, .
\label{44fivt}
\ee
In this case the $(2,2)$  lattice sum is
$\Gamma_{2,2} \ar{H^{\rm f}, h_1 - h_2}{G^{\rm f}, g_1 - g_2}$; the
shift
corresponds to the projection
$(-)^{m_2 G^{\rm f} + n_2(g_1-g_2)}$.

Finally, the computation of $B_2$ can be performed, and we obtain the
same results as in the $N_V=N_H=8$ model, summarized in Eqs.
(\ref{B288}) and (\ref{B288t}).
\vskip 0.3cm
\noindent{\sl c) The $N_V=N_H=2$ heterotic model}

\noindent
Here, we introduce the characters
\be
{\widehat{F}}_{1}\ar{\g,h}{\d, g} \equiv {1\over {\eta}^2}\,
{\vartheta}^{1/2}
\ar{\g-h_1-h_2-h_3}{\d-g_1-g_2-g_3}\,
{\vartheta}^{1/2}
\ar{\g+h_3}{\d+g_3}\,
{\vartheta}^{1/2}
\ar{\g+h_3-h_1}{\d+g_3-g_1}\,
{\vartheta}^{1/2}
\ar{\g+h_2-h_3}{\d+g_2-g_3}
\ee
and
\be
{\widehat{F}}_{2}\ar{\g, h}{\d, g} \equiv {1\over {\eta}^2}\,
{\vartheta}^{1/2}
\ar{\g}{\d}\,
{\vartheta}^{1/2}
\ar{\g+h_1-h_2}{\d+g_1-g_2}\,
{\vartheta}^{1/2}
\ar{\g+h_1}{\d+g_1}\,
{\vartheta}^{1/2}
\ar{\g+h_2}{\d+g_2}\, .
\ee
Note that the product
${\widehat{F}}_{1}\ar{\g,h}{\d, g}\,
{\widehat{F}}_{2}\ar{\g, h}{\d, g}$ has the same modular properties
as  $F_{1}\ar{\g,h}{\d, g}\, F_{2}\ar{\g, h}{\d, g}$. For the
model at hand,
\be
\Phi^V\ar{\g,\vec h}{\d,\vec g}={1\over \eta^2}\,
\vartheta \ar{\g}{\d}\,
\vartheta \ar{\g+h_1-h_2}{\d+g_1-g_2}
\label{22fiv}
\ee
(there is no dependence on $(H^{\rm f},G^{\rm f})$ and
$\vec h,\vec g$ stand again for
$(h_1,\ldots,h_5), (g_1,\ldots,g_5)$), and
\ba
\Phi^V\ar{\g,\vec h}{\d,\vec g}\,
\Phi^H\ar{H^{\rm f},\g, \vec h}{G^{\rm f},\, \d,\vec g}&=&
{\widehat{F}}_1\ar{\g+h_4+h_5,h}{\d+g_4+g_5,g}\,
F_1\ar{\g+h_4+h_5+H^{\rm f},h}{\d+g_4+g_5+G^{\rm f},g}\nn \\
&& \times
{\widehat{F}}_2\ar{\g,h}{\d,g}\,
F_2\ar{\g,h}{\d,g}\nn \\
&& \times 
F_1\ar{\g+h_5,h}{\d+g_5,g}\,
F_1\ar{\g+h_4+H^{\rm f},h}{\d+g_4+G^{\rm f},g}\nn \\
&& \times
F_2\ar{\g+h_5,h}{\d+g_5,g}\,
F_2\ar{\g+h_4,h}{\d+g_4,g}\, .
\label{22fivh}
\ea
The structure of $\Phi^V$ now shows that $G=U(1)^2$ and therefore
$N_V=2$.
The shifts in the $(2,2)$ and $(4,4)$ lattices are the same as those
in the
$N_V=4$ model.

An alternative embedding of $(H^{\rm f},G^{\rm f})$
is realized as follows: 
\be
{\tilde \Phi}^V\ar{H^{\rm f},\g,\vec h}{G^{\rm f},\, \d, \vec
g}={1\over
\eta^2}\,
\vartheta \ar{\g +H^{\rm f}}{\d +G^{\rm f}}\,
\vartheta \ar{\g+h_1-h_2+H^{\rm f}}{\d+g_1-g_2+G^{\rm f}}
\label{22fivt}
\ee
and
\ba
{\tilde \Phi}^V\ar{H^{\rm f},\g,\vec h}{G^{\rm f},\, \d,\vec g}\,
{\tilde \Phi}^H\ar{H^{\rm f},\g,\vec h}{G^{\rm f},\, \d,\vec g}&=&
{\widehat{F}}_1\ar{\g+h_4+h_5,h}{\d+g_4+g_5,g}\,
F_1\ar{\g+h_4+h_5+H^{\rm f},h}{\d+g_4+g_5+G^{\rm f},g}\nn \\
&&\times
{\widehat{F}}_2\ar{\g+H^{\rm f},h}{\d+G^{\rm f},g}\,
F_2\ar{\g+H^{\rm f},h}{\d+G^{\rm f},g} \nn \\
&&\times
F_1\ar{\g+h_5,h}{\d+g_5,g}\,
F_1\ar{\g+h_4,h}{\d+g_4,g}\nn \\
&&\times
F_2\ar{\g+h_5,h}{\d+g_5,g}\,
F_2\ar{\g+h_4,h}{\d+g_4,g}\, ;
\label{22fivht}
\ea
this also leads to $N_V=N_H=2$.
As for the $N_V=4$ model defined through (\ref{44fih}) and
(\ref{44fivt}), it is necessary to perform a double shift in the
$(2,2)$ lattice with the projection
$(-)^{m_2 G^{\rm f}+n_2 (g_1-g_2)}$.
The shift in the $(4,4)$ lattice is identical to the one
used together with the above $\Phi^V$
defined in (\ref{22fiv}).

Here also the second helicity supertrace is given in
(\ref{B288}) or (\ref{B288t}).

\subsection{Some general comments on the heterotic duals}

Our first observation is that all models with $N_V=N_H$ defined above
have identical
$B_2$ helicity supertrace, which is given either by (\ref{B288}) or
by (\ref{B288t}), depending on the embedding of the $Z_2^{(\rm f)}$.
This universality follows from (\romannumeral1) the
modular-transformation properties
of $\Gamma_{2,2}^{\lambda}\ar{H^{\rm f}}{G^{\rm f}}$,
(\romannumeral2)
the condition $N_V=N_H$, and (\romannumeral3) the  spontaneous
breaking
of $N=4$ supersymmetry  to $N=2$. These three requirements fix
uniquely the functions $\Omega$. We thus expect
that this universality of $B_2$ will remain valid for all $N_V=N_H$
models in  which the $U(1)^r$ gauge group is extended to a larger
gauge
group $G_r$ with the same rank.

This can be checked explicitly in several examples, for instance the
model
with a gauge group
\be
G_r =SU(2)_{k=2}^8
\label{su2}
\ee
defined with the choice ($\vec h \equiv (h_1, h_2, h_3)=h$,
and similarly for $\vec g$)
\be
\Phi^V \ar{\g, \vec h}{\d, \vec g} =
F_1^3\ar{\g,h}{\d, g} \, F_2^3\ar{\g, h}{\d, g}
\ee
\be
\Phi^H\ar{H^{\rm f},\g , \vec h}{G^{\rm f}, \, \d , \vec g}=
F_1\ar{\g+H^{\rm f},h}{\d+G^{\rm f},g}
F_2\ar{\g +H^{\rm f},h}{\d +G^{\rm f},g}\, .
\ee
 From this model one obtains the model
$N_V=8$ by switching on discrete Wilson lines,
which break the $SU(2)_{k=2}$ factors to eight $U(1)$'s. Actually,
when $(h_4,g_4)$ is
turned on in Eq. (\ref{88fih}), we obtain the model $N_V=8$;
when it is
turned off we obtain instead (\ref{su2}). Switching on a
non-zero  $(h_4,g_4)$ is equivalent to switching on a Wilson line
defined as $Y_I=(2p_I+1)h_4+2q_I$, with $(p_I,q_I)$ integers.
Another model with $r=8$ is the one constructed in \cite{fhsv}, based
on $E_8$ at level
two. In that case $h$ defines discrete Wilson lines that break
$(E_8)_{k=2}$ to $SU(2)_{k=2}^8$.

The connection of our previous construction to $(E_8)_{k=2}$ and
$SU(2)^8_{k=2}$ is useful and describes
well  the action of   $Z_2^{(\rm f)}$  on the
heterotic and type II duals.
At the orbifold point of $K3\sim T^4/Z^{(\rm o)}_2$, the
intersection matrix of $H^2(K3)$ is given by the following elements:
\be
L_{IJ} = \left( \Gamma_8^{(\rm o)} \oplus \Gamma_8^{(\rm o)}
\oplus \sigma^1 \oplus
\sigma^1
\oplus \sigma^1 \right)_{IJ} \ ,\ \ \sigma^1= \left( \begin{array}{cc}
0 & 1 \\
1 & 0 \end{array} \right)\, ,
\label{lij}
\ee
with
\be
\Gamma_8^{(\rm o)} \oplus \Gamma_8^{(\rm o)} = (-A_1) \oplus
\cdots \oplus (-A_1)\, ,
\label{a1}
\ee
$A_1$ being the Cartan matrix of $SU(2)$.
On type IIA, $Z_2^{({\rm f})}$ has 8
positive and 8 negative eigenvalues in the
submatrix (\ref{a1}), and thus removes 8 of the 16 fixed points.

On the heterotic side, $Z_2^{(\rm f)}$ interchanges
the eight $SU(2)^{(1)}$ with the eight $SU(2)^{(2)}$.
This results in a level-two realization of  $SU(2)^8$.
Furthermore, on type IIA,
$Z_2^{({\rm f})}$ has 2 positive and 4 negative eigenvalues
on the torus intersection matrix $\sigma^1 \oplus \sigma^1 \oplus
\sigma^1$; in our constructions, it rotates by $\pi$
two $\sigma^1$ matrices $(\sigma^1 \to -\sigma^1)$
and leaves invariant the third $\sigma^1$.
On the heterotic side, $Z_2^{(\rm f)}$ acts as a $\pi$ rotation in
$T^4$ coordinates and as a translation in $T^2$ ones.
The 4 additional negative eigenvalues required by modular
invariance on the heterotic side can be embedded either in the gauge
group (leading therefore to $\Gamma_{2,2}^{\lambda=0}$) or directly
in the
$\Gamma_{2,2}$, where they produce
an asymmetric shift ($\Gamma_{2,2}^{\lambda=1}$).

In the models $N_V=4$ and 2, also
the actions of $Z_2^{D}$ on heterotic side and
on the intersection matrix of $T^4/Z_2$ correspond.
On type IIA,
the $Z_2^{D}$ projection has 8 positive and 8 negative
eigenvalues on the 16 twist fields; it acts by
exchanging eight $A_1$ matrices with eight others. Four such
exchanges are common also to $Z_2^{(\rm f)}$, and also four
are the exchanges that are common to $Z_2^{D}$ and $Z_2^{D'}$.

On the heterotic side, the corresponding operations act
as a lattice exchange on the $SU(2)^{16}$ gauge group and
as a translation in the six-dimensional internal space.
Since for each one of them the orbifold twist has 8
negative eigenvalues, modular invariance forces the shift to be
symmetric. On the compact space, therefore, the action of the
$Z_2^{D}$ projections is analogous to that of the $Z_2^{i}$, $i=1,\ldots,5$,
which break the gauge group.
When combined with the shift due to
$Z_2^{(\rm f)}$, we therefore obtain the same two different
embeddings of the shift in $\Gamma_{2,2}$ as for the $N_V=8$ model.

It is interesting to observe that the construction of Ref. \cite{fhsv}
is based on a different separation of $\Gamma_{2,2+8}$:
\be
\Gamma_{2,2+8}(T,U, \vec Y )=
\Gamma^{\lambda=1}_{1,1} (R) \ar{H^{\rm f}}{G^{\rm
f}}\, \Gamma^{+}_{1,1+8}\ar{H^{\rm f}}{G^{\rm f}}\, ,
\ee
where $\Gamma^{+}_{1,1+8}$ is a lattice invariant
under the  interchange of $\Gamma^{1}_{1,1+8}$ with
$\Gamma^{2}_{1,1+8}$ induced by $Z_2^{(\rm f)}$.
In this case, not only a rank 8 part but also one of the $U(1)$'s
of the compact space can be  enhanced to $SU(2)_{k=2}$.
This separation, in which only one circle is singled out,
is not useful for our purpose.
The separation of a shifted $\Gamma_{2,2}$ lattice is necessary
because we want to identify the perturbative  heterotic
moduli $T,U$ with the  moduli $T^2, T^3$ of  type IIA. This has led
us to the two heterotic constructions, based on $\Phi^V$ and
$\tilde \Phi^V$, where the part of the gauge group
that comes from the separated two-torus is always realized at the level
one.

The two classes of $N_V=N_H$ heterotic orbifolds, $\Phi^V$ and
${\tilde\Phi}^V$, 
correspond to  different regions in the moduli space
of the lattice $\Gamma_{2,2+r}\left( T,U,\vec Y \right)$. 
Here, we would like to argue that the perturbative constructions
based on ${\tilde\Phi}^V$ are the most suitable for  comparison with
the type IIA orbifolds presented in Section \ref{II}. 
In fact, in models based on $\Phi^V$, even though the $Z_2^{(\rm
f)}$ translation in the two-torus produces a
spontaneous breaking of the $N=4$ to $N=2$ supersymmetry,
{\it it does not
reproduce the situation of the perturbative type IIA orbifold}.
On type IIA, the restoration of the $N=4$ supersymmetry
is achieved by taking only appropriate limits
in the perturbative moduli: for the specific
lattice shifts we considered, the $N=4$ supersymmetry is restored
when the moduli $T^1$ and $T^2$ are large, while in the opposite
limit
supersymmetry is broken to $N=2$ as in a non-freely-acting orbifold.
In the heterotic construction based on $\Phi^V $, however,
the $N=4$ supersymmetry is always restored
in any decompactification limit, because, for any choice of
direction of the shift,
a ($\lambda=1$)-shifted lattice sum vanishes both
for large and for small moduli. It is therefore impossible
to find a translation that reproduces the perturbative properties of
the type IIA duals. In this case the map between the heterotic moduli
$T$ and $U$ and the type IIA moduli $T^2$ and $T^3$ is non-linear.
On the other hand, in the constructions with $\tilde{\Phi}^V$,
the $N=4$ supersymmetry is again spontaneously broken because of the
$Z_2^{(\rm f)}$ translation on the  $T^2$.
Here, however, 
the appropriate limit of $N=4$ supersymmetry restoration is
determined by the choice of the  shifts in  the
$\Gamma^{\lambda=0}_{2,2}$
lattice since
the term with $\Gamma^{\lambda=1}_{2,2}$ becomes irrelevant:
this kind of shifted lattice sum
vanishes in any decompactification limit.
For the particular $Z_2^{(\rm f)}$ shift we have considered,
$(-)^{m_2G^{\rm f}}$, in all the models based on $\tilde{\Phi}^V$,
the mass of the two extra gravitinos is
\be
m^2_{3/2} = {1 \over 4 \Im T \Im U }\, .
\ee
The $N=4$ supersymmetry is restored only
when  $R_2= \sqrt{\Im T \Im U} $ is large, whereas for small values
of the $(T,U)$ moduli we recover a genuine $N=2$ non-freely-acting
orbifold. This is precisely the perturbative behaviour of the above
type IIA dual orbifolds.

Finally, we would like to comment on another important issue, namely
on the appearance of extra massless states along submanifolds of the
moduli space. In particular we are interested in 
``$N=2$ singularities'' corresponding to enhancements where
$\Delta N_V-\Delta N_H \neq 0$. Perturbatively, in the type IIA
models of Section \ref{II}, there are no such singularities
in the $(T^2,T^3)$ plane
(as we already pointed out, $B_2=B_2\vert_{\rm massless}=N_V-N_H\equiv 0$).

On the heterotic side, however, already at the perturbative level, the 
massless part of the helicity supertrace $B_2(T,U)$ jumps 
across several submanifolds of the moduli space $(T,U)$.
This is a
straightforward consequence of the appearance of specific powers of
$\bar q$ in the $\Gamma_{2,2}^{\lambda} \ar{h}{g}$'s.
A detailed analysis of the rational behaviours
that appear in shifted lattice sums is given in
\cite{kkprn}. 
In the constructions based on $\Phi^V$
there are lines where the $U(1)^2$
gauge symmetry corresponding to the vectors originating from the
two-torus is enhanced to $SU(2) \times U(1)$, $SU(2) \times SU(2)$,
or  $SU(3)$. Furthermore, along some lines,  
16 or 22 extra hypermultiplets appear; most of these new hypermultiplets
are charged under the gauge group of the two-torus.
On the other hand, the only $N=4$ enhancements come from
the rank-8 part of the gauge group, while the part of the gauge group
that comes from the two-torus is always realized at the level one.
The models based on $\tilde\Phi^V$ have a similar behaviour,
although, depending on the choice of shift vectors, some of the would-be
$N=2$ singularities occurring along several submanifolds of the
$(T,U)$ plane cancel between the $\lambda=0$ and $\lambda=1$ term.
In fact, a careful analysis of the
helicity supertrace 
$B_2 \left(\tilde{\Phi}^V \right)$ (Eq. (\ref{B288t})) shows that
the $N=2$ singularities present in this class of models are
$SU(2)$ enhancements of one or both of the
$U(1)$'s of the torus and the appearance of new
hypermultiplets, all charged under the gauge group of the two-torus.
Also in these constructions, the part of the gauge group
that comes from the two-torus is always realized at the level one.

Despite the presence of the above $N=2$ singularities,
we expect that the heterotic amplitude corresponding
to the $R^2$ coupling analysed in the type II dual models should not be
sensitive to the perturbative enhancement of the massless spectrum. In
the next section we will see how this heterotic amplitude can indeed
be computed, by demanding regularity in the $(T,U)$ space,
making it therefore possible to establish the mapping
between the heterotic and the type II moduli.

\section{Heterotic gravitational corrections and a
test of duality}
\label{hetthr}

Having the explicit expression of $B_{2}({\tilde \Phi}^V)$
in terms of
$\Gamma_{2,2}^{\lambda}\ar{H^{\rm f}}{G^{\rm f}}$ and using the
techniques developed in Refs.
\cite{ 6auth, kkprn, dkl, mast, lest, hmbps},
we can calculate, on the
heterotic side, the perturbative gravitational and gauge corrections
in terms  of the  moduli $T_{\rm Het}$ and $U_{\rm Het}$ of the $(2,2)$
lattice. These must be identical to the 
analogous corrections in type IIA given in
terms of the type II moduli $T^2$ and $T^3$ (see Eqs. (\ref{thrint})
and (\ref{limn2}) for the gravitational coupling).

In order to compare the perturbative heterotic and type
II results, it is necessary to look
for an operator on the heterotic side that allows for the computation
of the same physical quantity as on the type II side. 
The usual
gravitational operator is given by  
\be
Q_{\rm grav}^2\equiv  Q^2\,
\bP_{\rm grav}^2\, , 
\label{Qgrav}
\ee
where $Q$ stands again for the left-helicity operator, and $\bP_{\rm grav}^2$, 
when inserted inside the one-loop vacuum amplitude, acts as 
${-1\over 2 \pi i}{\partial \over
\partial{\bar \tau}}$ on ${1\over \im \, \bar{\eta}^2}$;
namely, it acts on the contribution of the two right-moving transverse
space-time coordinates $\overline{X}^{\mu=3,4}$, {\it including their
zero-modes.} This latter fact is responsible for the appearance of 
a non-holomorphic gravity-backreaction
contribution, which ensures modular covariance but has no type II
counterpart. Indeed, the one-loop amplitude of the above operator is
\cite{infra, delgrav, hmbps, agn, gauge} \footnote{In general the
heterotic one-loop amplitude of an operator of the form $Q^2\, \bP^2$
reads:
$$
\left\langle Q^2\, \bP^2
\right\rangle_{\rm genus-one}
= \bP^2\,B_2\, ,
$$
where, in the l.h.s., $\bP^2$ acts as a differential operator on some
specific factor of $B_2$.}
\be
F_{\rm grav}\equiv \left\langle Q_{\rm grav}^2 
\right\rangle_{\rm genus-one}
= -{1\over 12}\,  
\left( \bE_2 -{3\over \pi\im}\right) B_2\, ,
\label{Fgrav}
\ee
where, in the class of models under consideration, 
$B_2$ is given in Eq. (\ref{B288t}) (or (\ref{B288}) in models
constructed with $\Phi^V$). The massless contribution to $F_{\rm
grav}$ is precisely the gravitational anomaly, $b_{\rm grav}= {24 - N_V
+N_H\over 12}$, which in all heterotic models at hand equals 2
($B_2\vert_{\rm massless}$ vanishes), at generic points of the $(T,U)$
moduli space.

The operator $Q_{\rm grav}^2$ is not suitable for comparison with the
type II result  (\romannumeral1) because of the non-holomorphic term it
generates; (\romannumeral2) because its amplitude is sensitive to the
$N=2$ singularities occurring in the $(T,U)$ plane: the gravitational
anomaly jumps along several rational lines where $N_V-N_H$ is no longer zero. 
We must therefore replace
$Q_{\rm grav}^2$ with an appropriate operator 
$Q_{\rm grav}^{\prime 2}=Q^2\, {\bP}^{\prime 2}_{\rm grav}$ 
such that  (\romannumeral1)  ${\bP}^{\prime 2}_{\rm grav}$ is 
essentially a universal combination of $\bP_{\rm grav}^2$
and some appropriate Cartan generators of the gauge group; 
(\romannumeral2) the massless
contribution of the corresponding amplitude should remain the
gravitational anomaly, and (\romannumeral3) this amplitude should be
regular everywhere in $(T,U)$, at least in the models constructed with 
${\tilde \Phi}^V$ .

To this purpose,
we introduce the following combinations of gravitational and gauge
operators (we only display their right-moving factor):
\ba
\bP^2_1 & \equiv & 12\,  \bP^2_{\rm grav}+\bP^2_{2,2}\, ;
\label{q1}
\\
\bP^2_2 & \equiv & 12\,  \bP^2_{\rm grav}+\bP^2_{2,2}
+{8\over N_V}\bP^2_{\rm gauge}\, .
\label{q2}
\ea
After insertion into the one-loop heterotic vacuum amplitude,
$\bP^2_{2,2}$ acts as ${-1 \over 2 \pi i}{\partial \over
\partial{\bar \tau}}$ on the modular-covariant factor of weight zero
$\im \,\Gamma_{2,2}^{\lambda}\ar{H^{\rm f}}{G^{\rm f}}$.
This amounts to inserting the sum of the two right-moving
lattice momenta ${\bar p}_1^2+{\bar p}_2^2$ of $T^2$, which correspond
to the Cartan of the $U(1)$ factor. The amplitude 
$\left\langle Q^2\, \bP^2_{2,2}
\right\rangle$ therefore generates the corresponding gauge-coupling
correction. To be more precise, we must in fact consider the integral  
$\int_{\cal F} {d^2 \tau \over \im}\left\langle Q^2\, \bP^2_{2,2}
\right\rangle \g (\tau,\bar \tau)$ ($\g (\tau,\bar \tau)$ is an
appropriate modular-invariant infrared-regularizing function
\cite{infra}). An integration by parts can be performed which leads
to vanishing boundary terms {\it all over the $(T,U)$ plane},
irrespectively of the specific behaviour of the lattice sums across
rational lines. This allows us to recast the above amplitude as:
\be
\bP^2_{2,2}\, B_2 \left(\tilde{\Phi}^V \right)=
\sumpf{1\over 2}
\sum_{\lambda=0,1}
\left(
\Gamma_{2,2}^{\lambda} \ar{H^{\rm f}}{G^{\rm f}} 
\left[
{1 \over 2 \pi i}{\partial \over
\partial{\bar \tau}}- {1\over 2\pi \im}
\right]
{\bOmega^{(\lambda)} \ar{H^{\rm f}}{G^{\rm f}}\over \bar{\eta}^{24}}
\right) 
\label{P22}
\ee
(in models constructed with ${\Phi}^V $, only a $\lambda =1$ term
appears). The differential operator inside the brackets is covariant since
${\bOmega^{(\lambda)} \ar{H^{\rm f}}{G^{\rm f}}\Big/ \bar{\eta}^{24}}$
has modular weight $-2$, but non-holomorphic as in the case of 
$\bP_{\rm grav}^2$.

Finally, $\bP_{\rm gauge}^2$ acts as
${-1\over 2 \pi i}{ \partial \over
\partial{\bar \tau}}$ on the ($\bar \tau $-covariantized) charge 
lattice of
the remaining right-moving gauge group $G_r$ of rank $r$,
namely on 
$(\im)^{N_V/2} \, \bar \eta^{N_V} \, {\overline{\Phi}}^V$
(or $\tilde{\overline{\Phi}}^V$). It is
modular-covariant since  the ${\Phi}^V$'s are of
weight zero, and contains a non-holomorphic $1/\im$ term. Notice that
in the
$U(1)^{N_V}$ models,
\be
\bP_{\rm gauge}^2 =\sum^{N_V}_{I=1}
\bP_I^2\, ,
\ee
where $\bP_I$ are the $U(1)$ zero-modes.  
After some straightforward algebra we obtain:
\ba
\bP^2_{\rm gauge}\, B_2 \left(\tilde{\Phi}^V \right)&=&
{N_V\over 48}\sumpf
\Gamma_{2,2}^{\lambda=0} \ar{H^{\rm f}}{G^{\rm f}} 
\left(\left(
\bE_2 -{3\over \pi\im}
-{1\over 2}\bH\ar{H^{\rm f}}{G^{\rm f}}\right)
{\bOmega^{(0)} \ar{H^{\rm f}}{G^{\rm f}}\over \bar{\eta}^{24}}
+96
\right)
\nn \\
&&+{N_V\over 48}\sumpf
\Gamma_{2,2}^{\lambda=1} \ar{H^{\rm f}}{G^{\rm f}} 
\left(
\bE_2 -{3\over \pi\im}
-{1\over 2}\bH\ar{H^{\rm f}}{G^{\rm f}}
\right)
{\bOmega^{(1)} \ar{H^{\rm f}}{G^{\rm f}}\over
\bar{\eta}^{24}}\, ,
\label{Pgauge}
\ea
where we have introduced the modular-covariant functions
\be
H{h\atopwithdelims[]g}={12 \over  \pi i} \partial_\t \log {\th
{1-h\atopwithdelims[]1-g}\over
\eta}
= \cases{\hphantom{-} \th_3^4 + \th_4^4  \sp (h,g)=(0,1)\cr
          -  \th_2^4 - \th_3^4  \sp (h,g)=(1,0)\cr
\hphantom{-} \th_2^4 - \th_4^4  \sp (h,g)=(1,1)\cr }
\label{H}
\ee
of weight $2$. Similarly, for the models based on ${\Phi}^V$, 
\be
\bP^2_{\rm gauge}\, B_2 \left(\Phi^V \right)=
{N_V\over 24}\sumpf
\Gamma_{2,2}^{\lambda=1} \ar{H^{\rm f}}{G^{\rm f}} 
\left(\left(
\bE_2 -{3\over \pi\im}
-{1\over 2}\bH\ar{H^{\rm f}}{G^{\rm f}}\right)
{\bOmega \ar{H^{\rm f}}{G^{\rm f}}\over \bar{\eta}^{24}}
+48\, \overline{\chi}\ar{H^{\rm f}}{G^{\rm f}}
\right)\, ,
\ee
where $\chi \ar{h}{g}$ is a modular-covariant form of weight zero:
\be
\chi{h\atopwithdelims[]g}=(-1)^{hg}
{\Omega^{(0)}\ar{h}{g}
\over
\Omega^{(1)}\ar{h}{g}}
= \cases{ {1\over 2} \frac{\th_3^4 + \th_4^4}{\th_3^2\,  \th_4^2}
\sp (h,g)=(0,1)\cr
          {1\over 2} \frac{\th_2^4 + \th_3^4}{\th_2^2\,  \th_3^2}
\sp (h,g)=(1,0)\cr
          {1\over 2} \frac{\th_2^4 - \th_4^4}{\th_2^2\,  \th_4^2}
\sp (h,g)=(1,1)\, .\cr }
\label{chi}
\ee

Here, we would like to pause and analyse several properties of the
various operators $P_A$  introduced so far
($P_{\rm grav}$, $P_1$, $P_2$,
$P_{2,2}$, $P_{\rm gauge}$, $P_I$).
The beta-function coefficients of such operators
are the constant term in the large-$\im$ expansion of
the corresponding genus-one amplitude:
\be
\bP^2_{A}\, B_2= b_A + {1\over {\bar q}}\,  \sum_{0 <p(\ell)
\ne 1}c_{\ell}\,
{\bar q}^{p(\ell)}\, .
\ee
Let us concentrate on the constructions relevant to our duality
purposes, namely those based on $\tilde{\Phi}^V$. At generic points in
the $(T,U)$ plane, $b\left(P_{2,2}\right)$ vanishes, and
therefore $b\left(P_{1}\right)=12\, b_{\rm grav}=24$. 
When a line of enhanced massless spectrum is approached, the
beta-function coefficients $b\left(P_{2,2}\right)$ and $b_{\rm grav}$
jump
in the opposite way: $\Delta b\left(P_{2,2}\right) =-12\, \Delta b_{\rm
grav}$. This implies that the combination (\ref{q1}) is smooth
along the entire $(T,U)$ moduli space and we have
$b\left(P_{1}\right)=24$ everywhere. On the other hand, in the models
under consideration, $b\left(P_{\rm gauge}\right)$ is the
diagonal sum of higher-level gauge beta-functions, and thus vanishes
without any discontinuity. As a consequence, 
$b\left(P_{2}\right)=12\, b_{\rm grav}=24$ in the whole $(T,U)$ plane.

Both operators $\bP_{1}^2$ and $\bP_{2}^2$ lead to the insertion of a
covariant derivative that contains a non-holomorphic $1/\im$ term.
There is a unique linear combination of these, which is purely
holomorphic and whose beta-function coefficient is 
$b^{\prime}_{\rm grav}\equiv
b\left(P^{\prime }_{\rm grav}\right)=
b^{\vphantom{\prime}}_{\rm grav}=2$
{\it everywhere in the $(T,U)$ plane}. This operator is
\ba
\bP^{\prime 2}_{\rm grav} &\equiv & 
-{1 \over 8} \bP_1^2 
+{5\over 24} \bP_2^2 
\nonumber \\
& = &
\bP^{2}_{\rm grav} +
{1\over 12} \bP_{2,2}^2+
{5 \over 3 N_V}\bP_{\rm gauge}^2 \, ,
\ea
and satisfies all the
requirements we have demanded at the beginning of the section. It
defines the modified gravitational operator for which the one-loop
amplitude $\left\langle Q_{\rm grav}^{\prime 2} 
\right\rangle$ reads:
\ba
\bP^{\prime 2}_{\rm grav}\, B_2 \left(\tilde{\Phi}^V \right)&=&
-{1\over 24}\sumpf
\Gamma_{2,2}^{\lambda=0} \ar{H^{\rm f}}{G^{\rm f}} 
\left(\left(
{1\over 6}\bE_2 -{1\over 2 \pi i}{\partial \over
\partial{\bar \tau}}
+{5\over 12}\bH\ar{H^{\rm f}}{G^{\rm f}}
\right)
{\bOmega^{(0)} \ar{H^{\rm f}}{G^{\rm f}}\over \bar{\eta}^{24}}
-80\right)
\nn \\
&&-{1\over 24}\sumpf
\Gamma_{2,2}^{\lambda=1} \ar{H^{\rm f}}{G^{\rm f}} 
\left(
{1\over 6}\bE_2 -{1\over 2 \pi i}{\partial \over
\partial{\bar \tau}}
+{5\over 12}\bH\ar{H^{\rm f}}{G^{\rm f}}
\right)
{\bOmega^{(1)} \ar{H^{\rm f}}{G^{\rm f}}\over \bar{\eta}^{24}}\, .
\label{Pprigra}
\ea

In the above expression, the contribution of the $\lambda = 1$ part
vanishes identically, while the action of the new  holomorphic
covariant derivative on  
${\bOmega^{(0)} \ar{H^{\rm f}}{G^{\rm f}}\Big/
\bar{\eta}^{24}}$ gives a constant. This fact is a direct consequence
of the identity
\be
{1\over 2 \pi i}\partial_{\tau} \chi \ar{h}{g}= 32\, (-1)^{hg}
{\eta^{24}
\over
\Omega^{(1)}\ar{h}{g}}\, .
\ee
We therefore obtain:
\be
\left\langle Q_{\rm grav}^{\prime 2} 
\right\rangle_{\rm genus-one}=
2\sumpf
\Gamma_{2,2}^{\lambda=0} \ar{H^{\rm f}}{G^{\rm f}} 
\, .
\ee
Finally, the perturbative heterotic gravitational corrections of the
models constructed with $\tilde{\Phi}^V$ are
\be
\Delta_{\rm Het}^{ N_V=N_H}=
2 \ifd
\left(\sumpf \Gamma_{2,2}^{\lambda=0} \ar{H^{\rm f}}{G^{\rm f}}-1
\right) \, .
\ee
For the specific case in which the shift $(H^{\rm f}, G^{\rm f})$
in $\Gamma_{2,2}$ is due to a translation of momenta, $(-1)^{m_2
G^{\rm f}}$, we get:
\be
\Delta_{\rm Het}^{N_V=N_H}= -
2  \log \Im T \left\vert \vartheta_4 \left( T \right)
\right\vert^4 -
2  \log \Im U \left \vert \vartheta_4 \left( U \right)
\right\vert^4  + \ {\rm const.} \,  
\label{dhet}
\ee

The heterotic models under consideration, based on $\tilde{\Phi}^V$,
were advertised to
be dual to the type II ground states presented in Section \ref{II}. It
is therefore important to observe that on the type II side, the
replacement of 
$Q^2_{\rm grav} $ by
$Q_{\rm grav}^{\prime 2}$
does not modify the perturbative results, Eq. (\ref{thrint}).
Indeed, owing to the absence of perturbative
Ramond--Ramond charges, the contribution of the duals of 
$\bP^2_{2,2}$ and
$\bP^2_{\rm gauge}$ is always zero, and the one-loop amplitude
$\left\langle Q_{\rm grav}^{\prime 2} \right\rangle_{\rm II}$ therefore
reduces to
$\left\langle Q_{\rm grav}^{2} \right\rangle_{\rm II}$.
We can now compare the perturbative gravitational corrections
in the heterotic and type II dual orbifolds. On type II, 
the full result
is computed at one loop and given by (\ref{thrint}).
Since in the large-moduli limit a linear term comes
only from the field $T^1$ (see Eq. (\ref{limn2})),
this modulus has to be identified with the
heterotic dilaton: the heterotic coupling receives
in fact a tree-level contribution, linear in the dilaton expectation
value. Including the one-loop computation (\ref{dhet}), we obtain:
\ba
{16 \, \pi^2 \over g^2_{\rm   grav}(\mu^{(\rm Het)})} & = & 16 \, \pi^2 \Im S 
-2 \log \Im T \left \vert \vartheta_4 \left( T \right) \right\vert^4
-2 \log \Im U \left \vert \vartheta_4 \left( U \right) \right\vert^4
\nn \\
&& +4 \log {M^{(\rm Het)} \over \mu^{(\rm Het)} } + \ {\rm const.} 
\, ,
\label{htr}
\ea
where $S$ is the heterotic axion--dilaton field and
\be
\Im S  ={1 \over g^2_{\rm Het} }\, .
\ee
In (\ref{htr}) we used the string scale $M^{(\rm Het)} \equiv 
{1 \over \sqrt{\alpha'_{\rm Het}}}$ and the infrared cut-off
of the heterotic string.

In order to match
the heterotic and type II results (\ref{limn2}) and (\ref{htr})
in the heterotic
weak coupling limit, $\Im S \to \infty$, it is necessary to identify
$T^1$ with $16\tau_S /N_V$
\footnote{We use the notation $\tau_S=4 \pi S$.}.
This map and its normalization are consistent with the interpretation
of the type II vacuum as an
orbifold limit of a $K3$ fibration. In the $N_V=8$ case the  base of
the fibration is   $T^2/Z_2 = {\bf P}^1$,
with, as volume form, half of that of the torus $T^2$;
the heterotic dilaton then corresponds to the volume form
of the base of the fibration, $\tau_S = T^1/2$.
In the case of $N_V=4$, the base of the fibration  is instead ${\bf
P}^1/Z_2^{D}$, where $Z_2^{D}$ acts on
the sphere ${\bf P}^1$ as a half-circumference translation, and
$\tau_S = T^1/4$.
Finally, when $N_V=2$ the base of the fibration is
${\bf P}^1/(Z_2^{D} \times Z_2^{D'})$ and $\tau_S = T^1/8$.

Comparison of Eqs. (\ref{thrint}) and  (\ref{htr}) is even more
suggestive. By identifying
$T$ with $T^2$  and
$U$ with $T^3$,  the perturbative corrections on the
heterotic side as a function of $T$ and $U$ are identical to the
corrections on the type II side as a function  of $T^2$ and $T^3$. These
identifications allow us to promote the large-$\Im S$ heterotic
corrections obtained so far to finite values
of $\Im S$. The type II result (\ref{thrint}) 
therefore provides the full, perturbative and non-perturbative,
correction. The sub-leading logarithmic dilaton dependence and the
infrared running are obtained from the expressions (\ref{thrint}), 
(\ref{limn2}) by substituting the type IIA string mass 
$M^{(\rm II)}$ and cut-off $\mu^{(\rm II)}$ with the
duality-invariant Planck mass $M_{\rm Planck}$ and the
physical cut-off $\mu$. By using the relations
\be
{M^{(\rm II)} \over \mu^{(\rm II)}}=
{M^{(\rm Het)} \over \mu^{(\rm Het)}}=
{M_{\rm Planck} \over \mu}~,
\ee
we can express the full dilaton dependence and infrared running
of the effective coupling constant
$16 \, \pi^2/g^2_{\rm grav}(\mu^{(\rm Het)})$, for the various models,
as:
\ba
&&-{3N_V \over 4} 
\log \left\vert \eta \left( {16 \tau_S \over N_V} \right)\right\vert^4
-\left(2-{N_V \over 4} \right) \log \left\vert \vartheta_4
\left( {16 \tau_S \over N_V} \right)\right\vert^4 \nn \\
&& -\left( {  B_4-B_2 \over 3} 
- 2b_{\rm grav} \right)
\log \Im \tau_S \,+\,
 {  B_4-B_2 \over 3}  
\log {M_{\rm Planck} \over \mu},
\label{inst}
\ea
where by $B_4-B_2$ we actually mean the constant, massless contribution
at a generic point in moduli space: the coefficients of
the two terms in the second line are given in Eq. (\ref{run}). 
By expanding the above expression as
\[4\pi\Im\tau_S-
\left(2+{N_V \over 2} \right)\log \Im \tau_S+ 2\, {\rm
Re}\,  \sum_{k}n_k\,  e^{i2 k\pi\tau_S }
\]
and comparing it with the corresponding expression valid for the
$N=4$ heterotic string compactified on $T^6$
($ \sim 2\, {\rm Re}\log |\eta(\tau_S )|$)
\cite{hmn=4, 6auth},
we can see that the instanton numbers $k$ are
restricted to integers which are  respectively multiples of
2 ($N_V=8$ and $N_V=4$) and 4 ($N_V=2$)
\footnote{This feature was already observed for the $N_V=8$ model,
in \cite{hmn=2}.}. 

The case of heterotic models based on $\Phi^V$ deserves several
remarks. The one-loop amplitude of the modified gravitational operator
$Q_{\rm grav}^{\prime 2}$ reads:
\be
\bP^{\prime 2}_{\rm grav}\, B_2 \left(\Phi^V \right)=
2
\sumpf
\Gamma^{\lambda=1}_{2,2} \ar{H^{\rm f}}{G^{\rm f}} \, 
\overline{\chi} \ar{H^{\rm f}}{G^{\rm f}}\, .
\label{Pgauge2}
\ee

Expression (\ref{Pgauge2}) can still be integrated to give the
thresholds (see
\cite{kkprn} for details on such integrals). However, the heterotic
result in this case does not match the perturbative gravitational
thresholds of the type II models. In particular, we observe $N=2$
singularities along lines where $\Delta b_{\rm grav}^{\prime}\neq 0$, since 
$\Delta b(P_{2,2})\neq -12\Delta b_{\rm grav}$ 
\footnote{This is due to the appearance of hypermultiplets
uncharged under the gauge group.},
while $b(P_{\rm gauge})$
remains zero everywhere in the $(T,U)$ plane. 
This means that in the case at hand the relation between the type
IIA and heterotic perturbative moduli is not simply $T^2=T$, $T^3=U$.
The same conclusion was drawn when analysing the restoration of $N=4$
supersymmetry with respect to the various decompactification limits.

The difference between the two classes of models constructed in
Section 3 lies in
the choice of the discrete Wilson lines. The models
that are based on $\tilde \Phi^V$ are the duals of the type IIA
constructions
of Section 2, with the identifications $\tau_S={N_V\over 16}T^1$,
$T=T^2$, $U=T^3$. The models based on $\Phi^V$ correspond instead to a
choice of discrete Wilson lines that do not correspond
to the type II constructions presented in Section 2. In
Ref. \cite{hmn=2}, assuming the heterotic/type II duality, the
authors made a proposal for
the full non-perturbative gravitational corrections in the  $N_V=8$
model, as a function of all the vector-multiplet moduli:
$\Delta_{\rm grav}(\tau_S,T,U,Y_1,\ldots,Y_8)$.
This proposal is  based on the uniqueness properties
of the automorphic forms of the Calabi--Yau threefold with
$h^{1,1}=h^{2,1}=11$. In order to test this through a
comparison of heterotic and type IIA strings, special values of
the Wilson lines $\vec Y$ must be chosen; this necessarily leads to the
explicit constructions we have considered, which allow
a test of the heterotic/type II duality conjecture.

\section{Comments}\label{con}

In this work  we explicitly constructed heterotic and type II
dual pairs; we verified the duality conjecture not only for
a model with $N_V=8$, but also for  models
with $N_V=4$ and 2. A relevant choice of Wilson lines $\vec Y$
on the heterotic side defines the $\tilde \Phi^V$ constructions. The
heterotic/type II duality is not verified for the gravitational
corrections, but for a modified gravitational and gauge combination
associated to the operator $Q^{\prime 2}_{\rm grav}$ introduced in
Section~4. This operator has the property that it is regular, without
singularities in the entire $(T,U)$ moduli space.
We found that {\it the type II corrections $\Delta_{\rm II}^{
N_V=N_H}(T^1,T^2,T^3)$ provide the
complete, perturbative and non-perturbative heterotic corrections}.
This remarkable  property is due to the universality of the
$N=2$ sector on heterotic side and of the
corresponding $N=(2,2)$ sectors on type II.
Indeed, the heterotic $N=2$ sector is universal and independent of
the
particular choice of Wilson lines $\vec Y$:
it is the same for all the $N_V=N_H$ models with a separation
of the  $\Gamma^{\lambda}_{2,2}\ar{H^{\rm f}}{G^{\rm
f}}$ lattice as in the $\tilde \Phi^V$ models.

The heterotic instanton corrections
$n_k\, e^{ik\pi\tau_S}$ are due
to the Euclidean five-brane wrapped on the six-dimensional internal
space;
they depend only on $\tau_S$ and not on the other
moduli. The explicit expressions for these corrections are given in
Eq. (\ref{inst}). The permitted integers $k$  depend on $N_V$ and the
multiplicity coefficients $n_k$ are fully determined from type IIA.
The Olive--Montonen duality group is a subgroup of
$SL(2,Z)_{\tau_S}$, which depends  on $N_V$: it is  $\Gamma(2)$
when $N_V=8$, $\Gamma(8)$ when $N_V=4$, and $\Gamma(16)$ when $N_V=2$.

\vskip 0.3cm
\centerline{\bf Acknowledgements}

\noindent
C. Kounnas
and A. Gregori thank the Centre de Physique Th\'eorique de l'Ecole
Polytechnique for hospitality.
P.M. Petropoulos thanks the CERN Theory Division
and  acknowledges financial
support from the EEC contract TMR-ERBFMRXCT96-0090.

\end{document}